\documentclass[twocolumn,twocolappendix]{openjournal}
\usepackage{graphics}
\usepackage{amssymb}
\usepackage{bbm}
\usepackage{amsmath,textcomp,array}
\usepackage{bm}
\usepackage{xcolor}
\usepackage{comment}
\usepackage{wrapfig}
\usepackage{natbib}
\usepackage{url}
\usepackage{hyperref}

\begin{document}

\title{Investigating The Effects of Early Dark Energy on Large-scale Structure Within the EDENS Suite}

\shorttitle{The Effects of EDE on LSS in the EDENS Suite}

\author{Sophie Hodgson$^{1,\star}$}
\author{Patrick Wells$^{1}$}
\author{Katrin Heitmann$^{1}$}
\author{Niyantri Krishnan$^{1}$}

\affiliation{$^1$High Energy Physics Division, Argonne National Laboratory, 9700 South Cass Avenue, Lemont, IL 60439, USA}

\thanks{$^{\star}$E-mail: sihodgson@ucdavis.edu}

\shortauthors{Hodgson et al.}

\begin{abstract}
Early Dark Energy (EDE) models have been suggested as a possible solution to the so-called Hubble Tension, a discrepancy of the measurement of the Hubble constant at early and late times. In this paper, we investigate the effects of EDE on large-scale structure probes of cosmology via the EDENS (Early Dark Energy N-body Simulations) suite. The EDENS suite extends the reach of available EDE simulations considerably by adding volume and resolution. We derive several key metrics, such as the halo mass function, the nonlinear power spectrum, and the concentration-mass relation. Furthermore, we implement a halo occupation distribution model to populate the simulations with synthetic galaxies. This allows us to measure the galaxy-galaxy correlation function and galaxy bias. We choose an EDE model that is consistent with observations across several probes
and allows for the increase of the present day value of the Hubble constant to resolve the Hubble tension. This model adds three more parameters to the standard $\Lambda$CDM model. Additionally, it requires small shifts in the best-fit $\Lambda$CDM cosmological parameters to accommodate existing observational constraints. We find significant differences between the standard $\Lambda$CDM model and the EDE model, suggesting that some of our chosen metrics may allow us to distinguish EDE from $\Lambda$CDM, and future observations should help further constrain possible cosmological models. We release outputs from our simulations via the OpenCosmo data portal.
\end{abstract}

\section{Introduction}

The $\Lambda$CDM concordance model of cosmology has been remarkably successful over the last several decades, even as the quantity and quality of available data has increased \citep[e.g.][]{Riechers_2022}. Despite this success, there are a number of discrepancies that suggest the model could be incomplete \citep{2022JHEAp..34...49A}. Most recently, measurements from the Dark Energy Spectroscopic Instrument (DESI), combined with cosmic microwave background (CMB) and supernova data, have shown hints of a preference for evolving dark energy, with parametric and non-parametric reconstruction favoring an equation of state that crosses the phantom divide $w=-1$~\citep{2025PhRvD.112h3511L}. Constraints from CMB+DESI  on the neutrino mass sum are pushed remarkably close to--and in some analyses below--the minimum mass sum implied by oscillations~\citep{2025PhRvD.112h3513E}. Intriguingly, recent work suggests that these findings may not represent two independent anomalies, but rather different manifestations of the same underlying geometric inconsistency between datasets (for a recent study, see, e.g.~\citealt{2026arXiv260318131W}). Upcoming data from cosmological surveys spanning different wavebands and cosmic times, together with a detailed understanding of possible systematics and careful theoretical considerations, will help in the future to shed light on these fundamental questions. 

In this paper, we focus on another pressing cosmological puzzle to be solved, the so-called ``Hubble Tension''. The Hubble tension refers to the discrepancy between values of $H_0$ measured from early universe and late universe probes. This discrepancy has become too large to ignore, with some work placing it as high as $6\sigma$ \citep[e.g.][]{Di_Valentino_2021, Cai_2026}. There are studies, however, that find no tension, or tensions only up to around $3\sigma$ \citep[e.g.][]{Hoyt_2026,freedman2025statusreportchicagocarnegiehubble}, creating disagreement over the severity and existence of the tension. While the debate over the Hubble tension is still ongoing, it remains an active field of research to explore possibilities for this discrepancy. These range from unknown systematics in one or more probes to entirely new physics \citep[see][for recent reviews]{Knox_2020, Cai_2026}. In this work, we examine an entry into the latter category known as Early Dark Energy (EDE). According to a recent comparison study of solutions to the Hubble tension, EDE models are still one of the most promising avenues to explore~\citep{2026arXiv260713282S,2026arXiv260713283S}.

Early dark energy models introduce an additional component to the early universe in the form of a scalar field that behaves like dark energy at very early times before falling off around recombination \citep[e.g.][]{poulin_2023}. This additional component increases the pre-recombination expansion rate, leading to a corresponding decrease in the sound horizon. This decrease leads to an increase in the inferred angular diameter distance to the surface of last scattering and, by extension, $H_0$. In addition, EDE models tend to favor a greater dark matter density and steeper tilt of primordial perturbation spectra when fit to data from the CMB \citep[e.g.][]{shen2024earlygalaxiesearlydark,poulin_2023,Smith_2020}. New data from the Atacama Cosmology Telescope (ACT) DR6~\citep{ACT_DR6} show that there is no preference for an extended model such as EDE over $\Lambda$CDM, yet the two models provide practically the same fit to observational data. In a recent paper, the impact of these new data in combination with data from DESI DR2~\citep{2025PhRvD.112h3515A} on the validity of EDE has been explored~\citep{2026PhRvD.113f3519P}. They found that EDE still has promise as a valid model of the universe and could provide a solution to the Hubble tension. For a more thorough overview of various EDE models and their cosmological implications see \citet{poulin_2023}.

In this work we focus on a particular realization of EDE, an oscillating scalar field axion model added around the time of matter-radiation equality. This model was studied in detail by \citet{Smith_2020}, who showed that it provides an excellent fit to measurements of the CMB, baryon acoustic oscillations (BAO), and supernovae luminosity distance measurements. Their analysis included data from \textit{Planck} 2015~\citep{2016A&A...594A..11P}, measurements from SH0ES~\citep{2019ApJ...876...85R}, several BAO measurements~\citep{2011MNRAS.416.3017B,2015MNRAS.449..835R,2017MNRAS.470.2617A} and the Pantheon3 supernovae dataset~\citep{2018ApJ...859..101S}. The model features a slow-roll potential motivated by ultra-light axions of the form 
\begin{equation}
    V_n(\phi)  = m^2f^2\left[ 1-\cos(\phi/f) \right]^n,
    \label{eq:potential}
\end{equation}
where $m$ is the mass of the field, $f$ is the axion decay constant, and $n$ is a power-law index. For $n=1$ this is a well-established axion potential; phenomenological consequences of higher powers of $n$ are discussed in detail in \citet{Smith_2020}. \citet{Smith_2020} obtain the equation of state by inserting the potential into the Klein-Gordon equation and solving the linearized version. They derive best-fit model parameters for a $\Lambda$CDM model, varying the Hubble parameter $H_0$, $\omega_b$, $\omega_{\rm cdm}$, the amplitude and slope of the primordial power spectrum ($A_s$, $n_s$), and $\tau_{\rm reio}$ and two EDE models, one with $n=3$ fixed and three additional EDE parameters ($z_c$, $f_{\rm EDE}$, $\Theta_i$) and one with $n$ as a fourth free EDE parameter (see Table~I in \citealt{Smith_2020}). In our paper, we focus on their results for the best-fit $\Lambda$CDM model and the case for an $n=3$ potential. Newer constraints including ACT DR6 and DESI DR2 measurements are still fully consistent with these parameters choices~\citep{2026PhRvD.113f3519P}.

EDE models have significant consequences for late-time structure formation probes which may be observable. Future large-scale surveys may result in more accurate and exhaustive measurements of medium to high redshift ($z\geq 1$) galaxies that could be compared to models such as the one studied in this paper. The differences between $\Lambda$CDM and EDE models generally increase with redshift \citep[e.g.][]{Klypin_2020}, making the high-redshift universe a prime target for constraining cosmological models. Structure formation simulations are therefore crucial for evaluating these EDE models. In this paper, we present a comparison of different structure formation measurements from two gravity-only simulations performed with the Hardware/Hybrid Accelerated Cosmology Code (HACC)~\citep{2016NewA...42...49H}. We compare a $\Lambda$CDM cosmology to an EDE cosmology, and measure several statistics: the halo mass function (HMF), the concentration-mass relation, and the nonlinear matter power spectrum. Furthermore, we implement a halo occupation distribution (HOD) model \citep{2000MNRAS.318.1144P,2000MNRAS.318..203S,2001ApJ...546...20S,2002ApJ...575..587B} to populate the simulations with synthetic galaxies to obtain 3D galaxy distributions, 2-point galaxy correlation functions, the nonlinear galaxy power spectra, and the galaxy bias. For each of these statistics, we analyze the differences between the two models.

As part of this paper, we publicly release a range of simulations outputs. The majority of the analysis was carried out for redshifts $z=0, \; 0.5,\; \text{and} \; 1$. As such we release these outputs, and in addition $z=0.1$, for public use on the web-based portal introduced in \citet{Heitmann_2019}\footnote{https://cosmology.alcf.anl.gov}, as well as on the OpenCosmo portal \citep{wells2026_opencosmo}\footnote{https://opencosmo.science}. 

The paper is organized as follows. In Section~\ref{sec:sims} we describe the simulations and discuss our parameter choices. In Section~\ref{sec:results} we present the results from the comparison of the two different models. We describe the data release products in Section~\ref{sec:data}. Finally, we present our conclusions and outlook in Section~\ref{sec:conclusions}.

\section{Simulations and Parameters}
\label{sec:sims}

\subsection{Simulation Set-up}

The results shown in the following sections draw from the EDENS suite (Early Dark Energy N-body Simulations), two N-body simulations run on the Perlmutter supercomputer\footnote{https://docs.nersc.gov/systems/perlmutter/architecture/} at NERSC\footnote{https://docs.nersc.gov/}. These simulations have been carried out with HACC, a cosmology code that has been developed for more than a decade to run on any supercomputing architecture available, from many-core to hardware-accelerated architectures~\citep{2012arXiv1211.4864H,2016NewA...42...49H}. Most recently, hydrodynamics capabilities were added to HACC, resulting in CRK-HACC \citep{2023ApJS..264...34F,2025hpcn.conf...25F}. For this paper, we employ the gravity-only version. HACC provides an extensive suite of analysis tools that are run on-the-fly to generate halo catalogs for different halo mass definitions and matter power spectra at many outputs.

Similar EDE simulations were presented in \citet{Klypin_2020}, where the best-fit parameter values from the $n=\text{free}$ potential in \citet{Smith_2020} were used for their EDE model. However, \citet{Klypin_2020} rely on particle-mesh (PM) simulations only, considerably restricting their force resolution. \cite{Murgia_2021} carried out simulations with Gadget-3 with higher force resolution. Our simulations cover larger volumes at higher mass resolution, enabling more detailed studies of structure formation. This higher resolution allows us to extend previous work by looking further into halo abundances and halo concentration measurements, and by implementing an HOD model to determine observational tracers such as the 2-pt correlation functions and galaxy power spectra. We enable a direct comparison between the two simulations by choosing the same initial phases, while \citet{Klypin_2020} rely on the generation of many realizations to reduce cosmic variance. With these differences, our studies are very complementary to previous work. 

Each simulation covers a (1200 Mpc)$^3$ volume and evolves 2048$^3$ particles. The mass of each particle in the $\Lambda$CDM simulation is $7.83\times 10^9 M_{\odot}$, while the EDE simulation has a particle mass of $8.55\times 10^9 M_{\odot}$. This mass resolution allows us to identify halos but is not sufficient to resolve smaller subhalos. Given that our main aim in this paper is to characterize statistics relevant for cosmological measurements and to populate the simulation with galaxies using approximate methods, the mass resolution is fully sufficient.

Due to its rapid decay after recombination and negligible contribution at late times, EDE does not directly affect the evolution of large-scale structure. Its impact is instead imprinted on the matter distribution at early times, both directly and through shifts in the standard cosmological parameters. We therefore incorporate EDE into our simulations through the initial conditions, after which the evolution proceeds as in a standard $\Lambda$CDM cosmology with the corresponding cosmological parameters.

Each simulation is evolved from $z_{in}=200$ to $z=0$ , with initial particle positions and velocities determined using the Zel'dovich approximation \citep{Zeldovich_1970}. Both simulations have the same initial phases to enable a direct comparison. As described in \citet{Heitmann_2021}, the reason for the high-redshift start is to allow for sufficient time for structures to develop at redshifts of interest. The initial transfer function input for each simulation was created using the publicly available CLASS code \citep{Diego_Blas_2011}. The initial particle positions and velocities are then created internally in HACC, after creating a realization of the initial matter power spectra from the appropriate transfer functions.

\subsection{Analysis Tools}
\label{analysis-tools}

Several methods for finding halos are readily implemented in HACC. For the given mass resolution in our simulations, halos down to about $10^{12}M_\odot$ are reliably captured. We choose the spherical overdenstiy (SOD) halo definition \citep[first described in][]{Press_and_Schechter}, using the mass definition to be 200 times the critical density. As this definition is used later in our HOD model, we employ $M_{200c}$ halos throughout this paper to maintain consistency within our analysis. We find similar results to those presented in Section~\ref{sec:results} when using a friends-of-friends (FOF) \citep[first introduced in][]{Davis_1985} approach to determining distinct halos. The center of the SOD halos are determined by an FOF halo-finder with a linking length of $b=0.168$. This method allows for efficient identification of SOD halos and generally avoids overlaps between halos. This procedure has been tested before in the literature, see, e.g., \citet{Heitmann_2021} and was found to produce results in excellent agreement with other SOD halo finders~\citep{2011MNRAS.415.2293K}.

HACC has a range of analysis tools that run on-the-fly as part of the CosmoTools analysis suite~\citep{2016NewA...42...49H}. Details about CosmoTools and the HACC on-the-fly analysis workflow can be found in~\cite{2019ApJS..245...16H}. Running the workflows in-situ has the major advantage that we do not need to store the raw particle files and that the analysis tools have access to the same large number of nodes the simulation itself is running on. CosmoTools has been optimized over the years and takes full advantage of the GPU-accelerated architecture of Perlmutter.

Once the simulation is complete, we convert the outputs to the OpenCosmo format and use the OpenCosmo Python toolkit\footnote{https://github.com/ArgonneCPAC/OpenCosmo} \citep{wells2026_opencosmo} for downstream data management and analysis tasks. Power spectra for halos and galaxies are computed using the pk package\footnote{https://github.com/AstroPatty/pk}, which computes the Fourier transform of the density field produced by depositing galaxies with a Cloud in Cell (CiC) mass deposition, with appropriate corrections for the window function and shot noise. 

\subsection{Parameter Choices}

We choose the cosmologies for the EDENS suite following \citet{Smith_2020}. Table~\ref{tab:param} lists all cosmological parameters for our two simulations. In addition, we list $S_8$, derived from the combination of $\Omega_m$ and $\sigma_8$. The EDE model features a high value of $H_0$ at $z=0$ to address the Hubble tension. To accommodate this value and still fit observations, the standard cosmological parameters (in particular $\Omega_m$ and $\sigma_8$) must be adjusted accordingly. This changes the values of standard $\Lambda$CDM parameters slightly in addition to the introduction of the three new EDE parameters ($f_{\text{EDE}},\: z_c,\: \text{and} \: \Theta_i$). The shift of the standard cosmological parameters can be straightforwardly accommodated. Several of these parameters are degenerate and the often quoted parameter $S_8=\sigma_8\sqrt{\Omega_m}/0.3$ for large-scale structure probes that combines $\sigma_8$ and $\Omega_m$ is -- not surprisingly -- still remarkably similar for both EDENS cosmologies. \citet{poulin_2023} provide a comprehensive discussion about constraints from different data sets and shifts in the best-fit cosmologies in the case of adding additional EDE parameters and also possible challenges for the models.

\begin{table}[h]
    \centering
    \begin{tabular}{c|c|c}
    \hline
         Parameter & $\Lambda$CDM & EDE \\
         \hline
         $H_0$ & 68.21 & 72.19 \\
         \hline
         $\omega_{b}$ & 0.0225 & 0.02253 \\
         \hline
         $\Omega_{\rm CDM}$ & 0.253 & 0.2506 \\
         \hline
         $10^9A_s$ & 2.216 & 2.215 \\
         \hline
         $n_s$ & 0.9686 & 0.9889 \\
         \hline
         $\sigma_{8}$ & 0.8281 & 0.8518 \\
         \hline
         $f_{\text{EDE}}(z_c)$ & - & 0.122 \\
         \hline
         $\Theta_{i}$ & - & 2.83 \\
         \hline
         $\log{10}(z_c)$ & - & 3.562 \\
        \hline
        \hline
        $S_8$ & 0.830 & 0.842 \\
        \hline
    \end{tabular}
    \caption{Various simulation parameters, with cosmologies taken from \citet{Smith_2020}. $z_c$ is the critical redshift at which EDE becomes dynamical. $f_{\text{EDE}}$ is the fraction of energy density contributed by EDE. $\Theta_i$ is the initial field density before oscillation. $S_8$ is a derived parameter.}
    \label{tab:param}
\end{table}

Analysis of the most recent ACT data release \citep{ACT_DR6} have shown that there is no preference for an EDE cosmology over $\Lambda$CDM. However, the two models produce almost the same fit to the data, which does not rule out EDE as a valid model for solving the Hubble tension, a result confirmed by \citet{2026PhRvD.113f3519P}. As such, we move forward with the \citet{Smith_2020} model as a possible solution to the tension and compare it to the best-fit $\Lambda$CDM cosmology. For each model we carry out one simulation. The choice of the same initial phases for both simulations enables a direct comparison without noise from cosmic variance.

\begin{figure}
    \centering
    \includegraphics[width=1\linewidth]{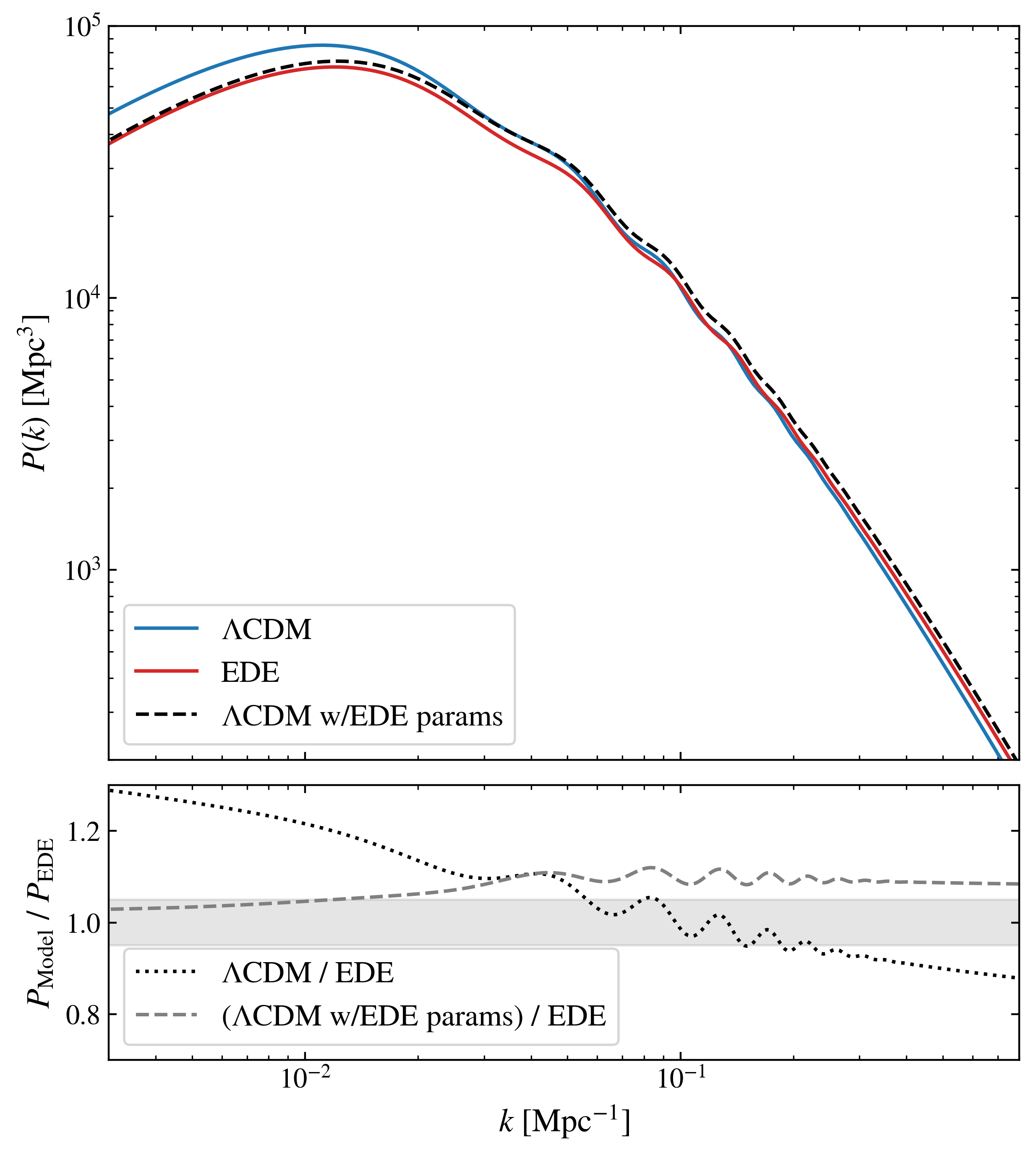}
    \caption{(Upper) Linear power spectra for the two models ($\Lambda$CDM in blue and EDE in red), and a $\Lambda$CDM model with the same parameters as the EDE model (black dashed). (Lower) The ratio of the two $\Lambda$CDM models to the EDE model. The 5\% range in either direction is shaded in grey.}
    \label{fig:CLASS_linear}
\end{figure}

To gain more insight in how the different cosmological parameters affect each model, we analyze the linear power spectra derived from CLASS. Figure~\ref{fig:CLASS_linear} shows the linear power spectra at $z=0$ for our EDE and $\Lambda$CDM models, in addition to a $\Lambda$CDM universe featuring the same standard parameters as our EDE model. The shift in the standard cosmological parameters impacts the linear power spectrum for the $\Lambda$CDM cosmology (dashed black line) considerably, in particular on large scales. However, the altered $\Lambda$CDM model is similar to the EDE model at most scales. This shows that the addition of the three extra EDE parameters do not cause dramatic differences in the linear power spectrum. Furthermore, around the BAO scale at $k\sim0.1$ Mpc$^{-1}$ the EDE and $\Lambda$CDM models are within 5\% of each other, showing that our EDE model does indeed match BAO observations. This analysis implies that many of the differences we report in Sec.~\ref{sec:results} are the result of the shifts in the standard cosmological parameters rather than the additional physics added by EDE.

\section{Results}
\label{sec:results}
In this section we describe the major findings from the analysis of the EDENS suite. For the primary simulation, we focus on the HMF, the concentration-mass relation, power spectra. We additionally populate the simulation with galaxies using an HOD model, from which we measure the 2-point galaxy correlation function and power spectrum. The results provide insights into the effects of viable EDE models on large-scale structure probes of cosmology. They deliver predictions for current and future observations and help identify the most impactful measurements to constrain EDE models.

\subsection{Halo Abundance}
\begin{figure}[t]
    \centering
    \includegraphics[width=1.0\linewidth]{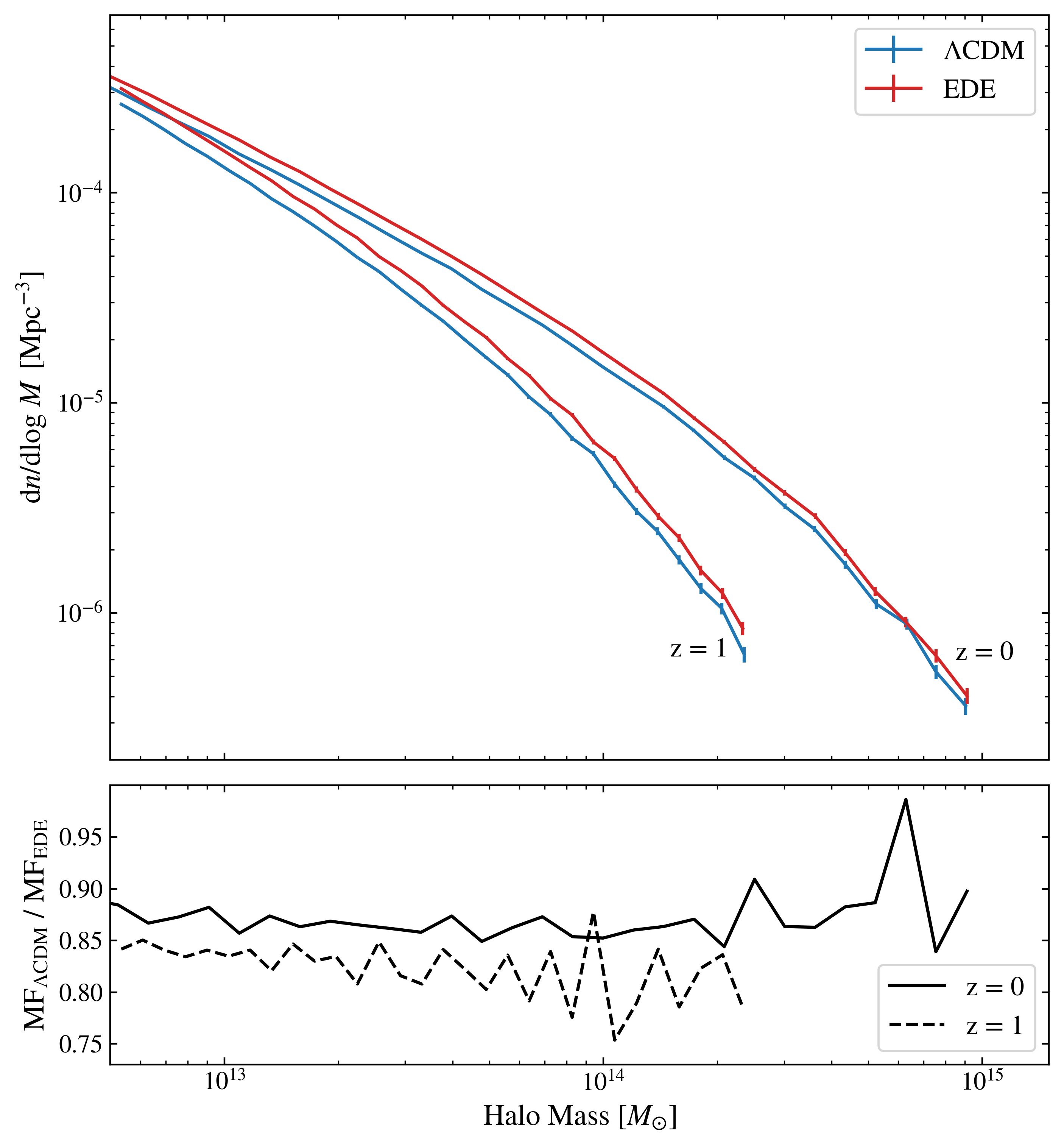}
    \caption{(Upper) Halo mass function (HMF) for $z=0$ and $z=1$ (labeled in black). $\Lambda$CDM is shown in blue, while EDE is shown in red. Error bars are given by a Poisson error. (Lower) The ratio of HMF$_{\Lambda\text{CDM}}$ to HMF$_{\text{EDE}}$. The solid line is $z=0$, while dashed is $z=1$.}
    \label{fig:HMF}
\end{figure}

Clusters of galaxies have long been recognized as an important source of cosmological information~\citep{2011ARA&A..49..409A}, and are are particularly sensitive to different dark energy models. Accurate predictions of the cluster mass function are based on measurements of the HMF from simulations, and as such we start our investigation with the HMF from the EDENS suite. We focus on the HMF as derived from SOD $M_{200c}$ halos. While our framework supports other halo mass definitions, we choose this halo mass definition since it is the input later into our HOD model. We therefore use consistent mass definitions throughout our investigations. Figure~\ref{fig:HMF} shows the stacked HMF for redshifts $z=0$ and $z=1$, for both simulations. Each measurement was obtained by binning the halos in equally-sized bins in log space, ensuring each bin contains at least 100 halos to reduce statistical error bars. A Poisson error is determined for each measurement point to obtain the error bars. The ratio of $\Lambda$CDM to EDE is shown in the lower panel of the figure. 

Figure~\ref{fig:HMF} clearly shows that the EDE simulation forms more, and more massive halos than the $\Lambda$CDM simulation. The values of the EDE HMF are higher by $\sim$12\% at $z=0$ across the entire mass range, and $\sim$18\% higher at $z=1$, again consistent between the low-mass and high-mass regimes. There is a slight decrease in the difference of the simulations in the high-mass regime (especially for $z=0$), however this is likely due to the small number of high mass halos and an artifact of our binning, rather than a significant effect from the cosmologies. The increased number of higher mass halos in the EDE simulation is likely due to the larger $\sigma_8$ value compared to the $\Lambda$CDM simulation. Despite EDE's increased expansion rate, a higher $\sigma_8$ value will lead to a higher clustering of mass, and thus more halos and/or more massive halos. The measurements for $z=1$ span a smaller range of halo masses as expected from large-scale structure growth over time. These results are in agreement with \citet{Klypin_2020}, where it was found that EDE forms 10\% more halos in the high mass regime. Similarly, the difference in the simulations grow as redshift increases, in agreement with findings in~\citet{Klypin_2020}. Furthermore, both \citet{shen2024earlygalaxiesearlydark} and \citet{EDE_galaxies_cosmic_rush} found the HMF of EDE to be higher than $\Lambda$CDM, and the differences to increase with redshift. While these authors utilized different models and higher redshifts than this work, our results generally match their conclusions. Therefore, the HMF may be a possible tracer of EDE, especially for higher redshifts, as the difference is greater than a few percent and may be detectable in observations from cosmological surveys carried out by, e.g., the Euclid and Rubin observatories \citep{Euclid,Ivezic_2019}.

\subsection{Concentration-Mass Relation}

Next we discuss our results for the concentration-mass relation. The concentration for each halo is obtained by fitting a Navarro-Frenk-White (NFW) profile \citep{1996ApJ...462..563N,1997ApJ...490..493N} of the form
\begin{equation}\label{eq:nfw}
\rho (r) = \frac{\delta \rho_{\rm c}}{(r/r_s)(1+r/r_s)^2},
\end{equation}
with $\delta$ being the characteristic dimensional density, and $r_s$ the scale radius of the NFW profile. The halo concentration is then simply determined by $c_\Delta=r_\Delta/r_s$, where $\Delta$ is the overdensity measured with the respect to the critical density. In this paper we choose $\Delta=200$. For a comprehensive discussion of the measurement of halo concentrations from N-body simulations, see, e.g., \cite{2018ApJ...859...55C}.

\begin{figure}
    \centering
    \includegraphics[width=1.0\linewidth]{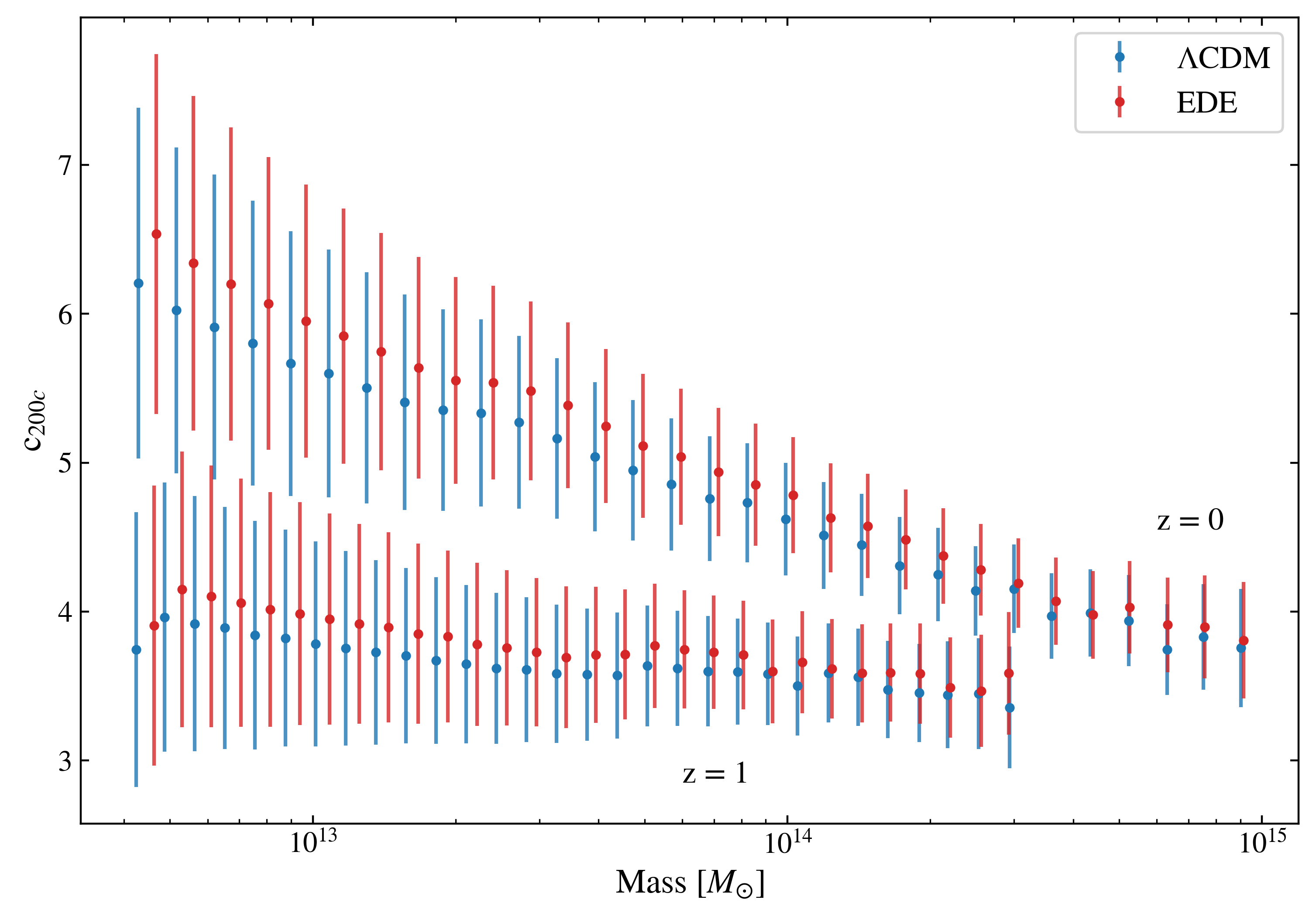}
    \caption{Concentration-mass relation for $\Lambda$CDM in blue and EDE in red. The upper curves are for $z=0$, with the lower curves corresponding to $z=1$. Error bars given by equation~\ref{cm_error}.}
    \label{fig:cM-relation}
\end{figure}

Figure~\ref{fig:cM-relation} shows measurements for both $z=0$ and $z=1$. The concentration of each halo is calculated in-situ by HACC, and can easily be binned similarly to the HMF. We create equal bins in log-space, taking the average concentration in each bin. The error bars are given by equation 19 in \citet{Heitmann_2021}:
\begin{equation}
\Delta c_i(M) = \sqrt{\frac{\Sigma_j \delta c_j}{N_i}+\frac{c^2(M)}{N_i}},
\label{cm_error}
\end{equation}
where $\delta c_j$ is the individual concentration error of each halo, and $\Delta c_i$ is the error for each bin. 

We find that the concentration of the halos in the EDE simulation are higher than those in the $\Lambda$CDM simulation. However, the two are very similar in the high-mass regime, and the error bars overlap significantly. We also find that the difference between the simulations decreases with redshift, showing an inverse relationship to that found in the HMF. The differences in the concentrations of the halos may be due, once again, to the higher $\sigma_8$ value used in the EDE simulation. Higher mass clustering will lead to higher concentration halos. 

\subsection{Matter Power Spectrum} 

Next we focus on the measurements of the nonlinear power spectrum from the simulations. Just like the HMF, the power spectrum holds important cosmological information. Most recently the Dark Energy Survey (DES) released cosmological constraints from galaxy clustering and weak lensing~\citep{2026arXiv260114559D}, relying on accurate predictions of the power spectrum from simulations and approximate methods calibrated with simulations. In the case of DES, the predictions were obtained from {\sc HMCODE2020}~\citep{2021MNRAS.502.1401M}, a halo model inspired approach. Other methods include Gaussian-process-based emulators~\citep{2023MNRAS.520.3443M} or, for less accurate but easy to use predictions, functional fitting forms~\citep{2012ApJ...761..152T}. 

\begin{figure}
    \centering
    \includegraphics[width=1\linewidth]{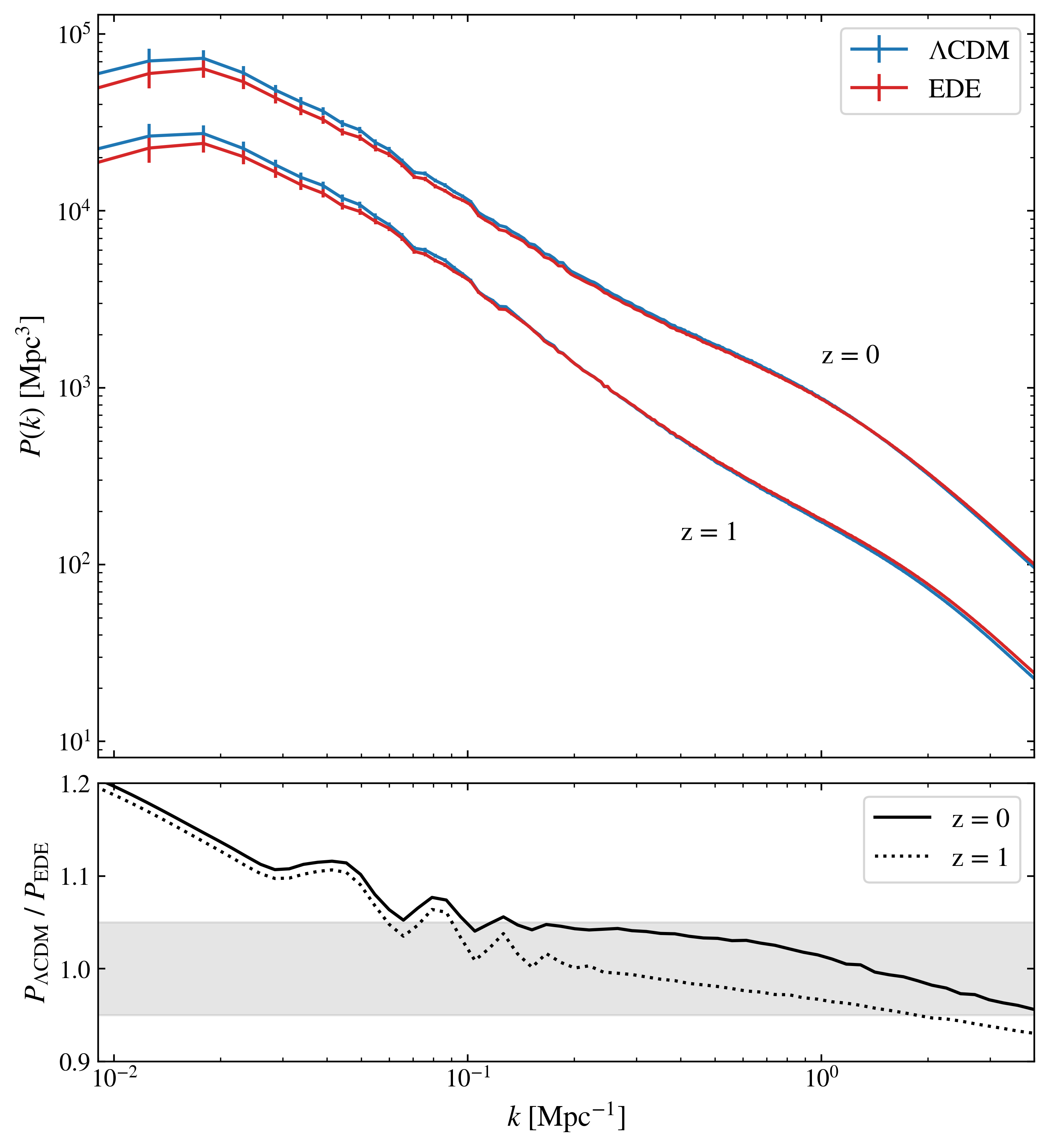}
    \caption{(Upper) Nonlinear matter power spectra. $\Lambda$CDM is shown in blue and EDE is shown in red. Error bars are given by $\sigma_{P (k_i)} = P(k_i)/\sqrt{N_{\rm nodes}(k_i)}$, assuming a Gaussian Random Field. The redshift is denoted with black text. The BAO wiggles are visible around $k=0.1\: \text{Mpc}^{-1}$. (Lower) Ratio of $P_{\Lambda \text{CDM}}(k)$ to $P_{\text{EDE}}(k)$ with $z=0$ corresponding to the solid line, and $z=1$ to the dashed line. The grey region represents a 5\% difference in either direction.}
    \label{fig:pk}
\end{figure}

For the EDENS suite, the nonlinear matter power spectrum is calculated on-the-fly at various checkpoints throughout the simulation runtime. We show results at two redshifts, $z=0,1$ in Fig.~\ref{fig:pk}.

The matter power spectra for both models are similar at small scales (high $k$), but differ on large scales (smaller $k$). The $\Lambda$CDM model has a higher power spectrum, until $k \sim 0.1 \: \text{Mpc}^{-1}$, where it starts to match the EDE model very closely, for both redshifts. These results are consistent with the earlier discussion of the linear power spectrum (Fig.~\ref{fig:CLASS_linear}). The differences at smaller $k$-values are likely due to the differences in the $n_s$ and $\sigma_8$ values between the simulations, as the EDE model has higher values than the $\Lambda$CDM model.  The BAO wiggles are visible in both redshifts around $k=0.1 \: \text{Mpc}^{-1}$. This matches the findings in \citet{Klypin_2020}, where the BAO wiggles were found around this wavenumber as well. Our results suggest a slightly higher contrast between the two models for redshift $z=0$, with our work showing a difference higher than 5\% around the BAO wiggles. Yet the remainder of our spectra match their results well, with our differences reaching similar levels to theirs at larger $k$ values. These differences are due to small differences in the cosmological models considered. Also note, that \citet{Klypin_2020} measure $k$ in $h$Mpc$^{-1}$ while we have chosen to remove $h$ from our units in the analysis to enable a cleaner comparison between the models.

\citet{Murgia_2021} present the power spectrum for an EDE cosmology and a best fit \textit{Planck} 2018 $\Lambda$CDM universe (once again in $h$Mpc$^{-1}$); they utilize different cosmological parameters for both EDE and $\Lambda$CDM than we implement, yet our results nearly match theirs. At $z=0$, around the BAO scale, their models have a difference of $\sim$5\%, which corroborates our results. We also see our difference decrease and flatten out after the BAO which is found in \citet{Murgia_2021} as well. We find agreement between this work and that of \citet{Murgia_2021} for $z=1$ as well. Our results indicate a smaller difference between the models with increasing redshift for most scales, which was also found in \citet{Murgia_2021}.

Our power spectra match previous findings in the literature and indicate the likelihood of this metric as a tracer for cosmological models. Our results suggest that the evolution of the BAO scale could be a tracer of EDE, especially for present day measurements, as the models have a difference close to 5\%.

\subsection{HOD Model}
Many recent explorations of EDE \citep[such as][which we extend here]{Klypin_2020} do not investigate its effects on galactic structures. As such, we generate a synthetic galaxy catalog using an HOD model to connect our simulations to observables and present more possible tracers for EDE within the current literature. At their most basic, HOD models calculate a likelihood function of a halo hosting a galaxy based on the mass of the halo, which is then sampled to populate the halos within the simulation with galaxies \citep{2000MNRAS.318.1144P}. HOD models are calibrated against specific sets of galaxies with certain properties (usually grouped into red or blue galaxies), making each particular HOD model valid only for a specific galaxy population. Each halo either hosts a ``central" galaxy, which resides at the center of the halo, or does not host any galaxies; this distinction is established by a halo mass threshold. A halo that hosts a central galaxy may then host ``satellite" galaxies, which are placed within the halo based on its density profile, and the number of which are determined again by halo mass. For a recent review of HOD models see \citet{Asgari_2023}. 

Luminous red galaxies (LRGs) are excellent tracers of the large scale structure of the universe \citep[e.g.][]{White_2011, Zhou_2023}. These galaxies reside in large halos (thus useful for our simulation resolution) and \citet{White_2011} have provided the required modeling parameters obtained from low redshift ($0.4 < z < 0.7$) BOSS observations. Other galaxy populations are harder to model, and as such are not included in this work. However, it should be noted that in reality there would be other types of galaxies present (besides LRGs), resulting in a larger total number of galaxies per halo than we have estimated here. While this model is relatively simple, with current literature pointing towards the effectiveness of extended models that include environmental factors \citep[first introduced in][]{Yuan_2021}, the \citet{White_2011} model provides an excellent baseline estimate for the distribution of LRGs within halos at redshift $z\sim0.5$. Our model is based on the best fit parameters in Table 2 of \citet{White_2011}, which have been listed in Table~\ref{tab:HOD} here for completeness. The HOD model is described by the following equations to calculate the probability of a halo of mass $M$ hosting $N$ central and satellite galaxies:

\begin{equation}
    N_{\text{cen}}(M) = \frac{1}{2} \: \text{erfc} \left[ \frac{\ln({M_{\text{cut}}/M})}{\sqrt{2} \: \sigma} \right],
\end{equation}

\begin{equation}
    N_{\text{sat}}(M) = N_{\text{cen}} \: \left( \frac{M-\kappa \: M_{\text{cut}}}{M_1} \right)^\alpha .
\end{equation}

$M_{\text{cut}}$ and $M_1$ are cutoff masses designed to ensure only sufficiently large halos host galaxies. $\sigma$, $\kappa$, and $\alpha$ are all fit parameters of the model and detailed in the model's original paper. The locations of the central galaxies are determined by the center of each host halo, while the locations of satellite galaxies are determined by an NFW profile, given in Eq.~\ref{eq:nfw}. 

\begin{table}
    \centering
    \begin{tabular}{c|c}
        lg$M_{\text{cut}}$ & 13.04 \\
        lg$M_1$ & 14.05 \\
        $\sigma$ & 0.94 \\
        $\kappa$ & 0.93 \\
        $\alpha$ & 0.97
    \end{tabular}
    \caption{The best fit parameters for the HOD model from \citet{White_2011}.}
    \label{tab:HOD}
\end{table}

\begin{figure*}[t]
    \centering
    \includegraphics[width=1\linewidth]{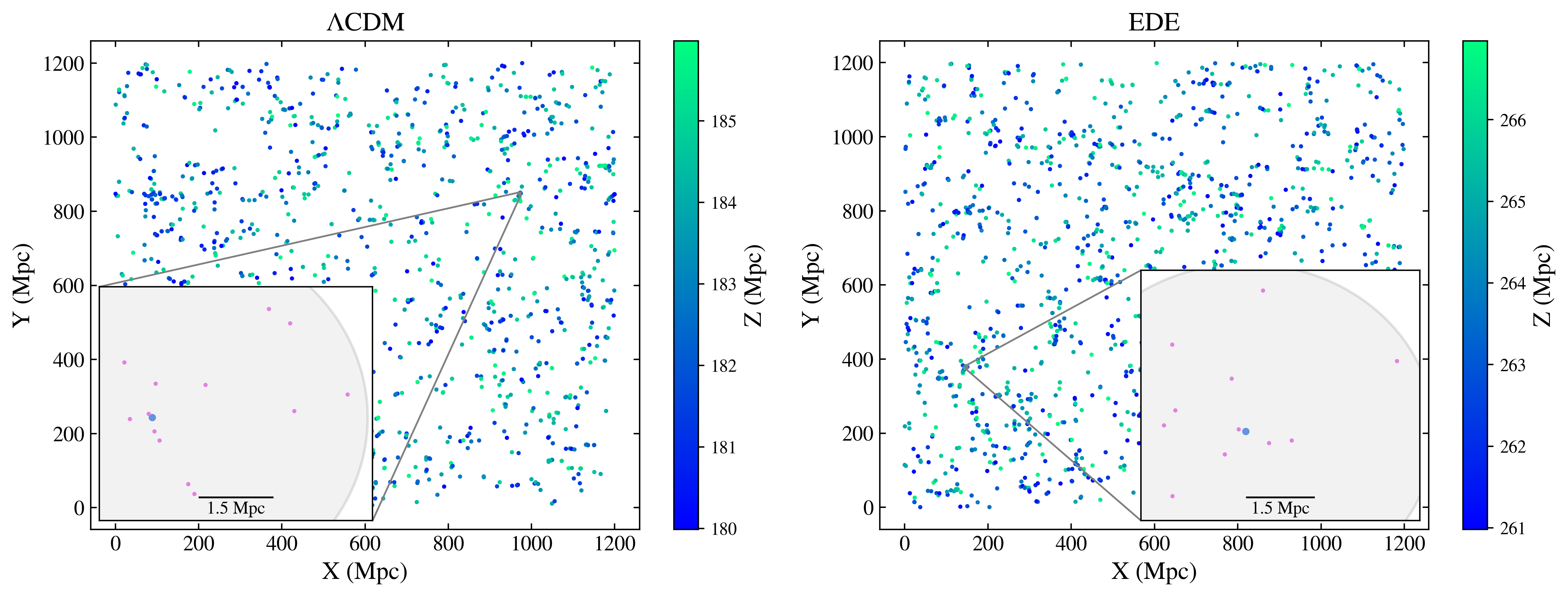}
    \caption{(Zoom) Specific halo of mass $\sim1.8\times 10^{15} M_{\odot}$ in each simulation. Centrals are shaded in blue, while satellites are shaded in pink. The host halo boundary is shaded in grey. (Body)  Distribution of synthetic galaxies within a 6 Mpc slice through the $z$-axis. Galaxies are shaded based on depth in the $z$-plane. The EDE model formed an average of 336575 galaxies while the $\Lambda$CDM model formed an average of 278803 galaxies over 150 iterations of the HOD model.}
    \label{fig:HOD-zoom}
\end{figure*}

We use the open source software HaloTools\footnote{https://github.com/astropy/halotools} \citep{Hearin_2017} v0.9.3 to implement the HOD model and create a synthetic galaxy catalog based on halo mass alone. We find that over 150 iterations of the model, the EDE simulation forms more galaxies than the $\Lambda$CDM simulation on average by about 21\% at redshift $z=0.5$. We visualize this distribution of the synthetic galaxies in Fig.~\ref{fig:HOD-zoom} and confirm the EDE simulation forms more galaxies. We plot the location of each galaxy, shaded with respect to depth in the $z$-axis. Figure~\ref{fig:HOD-zoom} shows a slice through the $z$-axis of the simulation box with a width of 6 Mpc. The zoomed-in region of the plot shows a comparison of two similar mass halos ($\sim 1.8\times10^{15} M_{\odot}$) in each simulation, showing that similar mass halos have similar numbers of satellite galaxies, as is expected from our HOD model. When binned by halo mass, the average number of satellites per halo differed by less than 1\% for nearly all bins between the two models. The galaxies are similarly clustered in both simulations, as shown by the scale bar in the zoomed-in plot. This is in agreement with the earlier discussions for the concentration-mass relation, where we found that while the EDE simulation generally produced more concentrated halos, the two models were very similar in the high-mass regime. The larger body of the figure shows the distribution of galaxies across the cross-sectional area of the simulation box, showcasing the difference in density of galaxies. Please note that the marker size of each galaxy does not correlate to physical size, and is simply denoting the location of each galaxy.

\subsection{Galaxy 2-pt Functions}

From our synthetic galaxy catalog, we calculate the 2-point galaxy-galaxy correlation function, using the TreeCorr\footnote{https://github.com/rmjarvis/TreeCorr} package \citep{Jarvis_2004}. The HOD implementation from \cite{White_2011} only allows us to study one redshift, $z=0.5$ and we therefore will not investigate the redshift evolution of the correlation function for the EDENS suite. 

\begin{figure}
    \centering
    \includegraphics[width=1\linewidth]{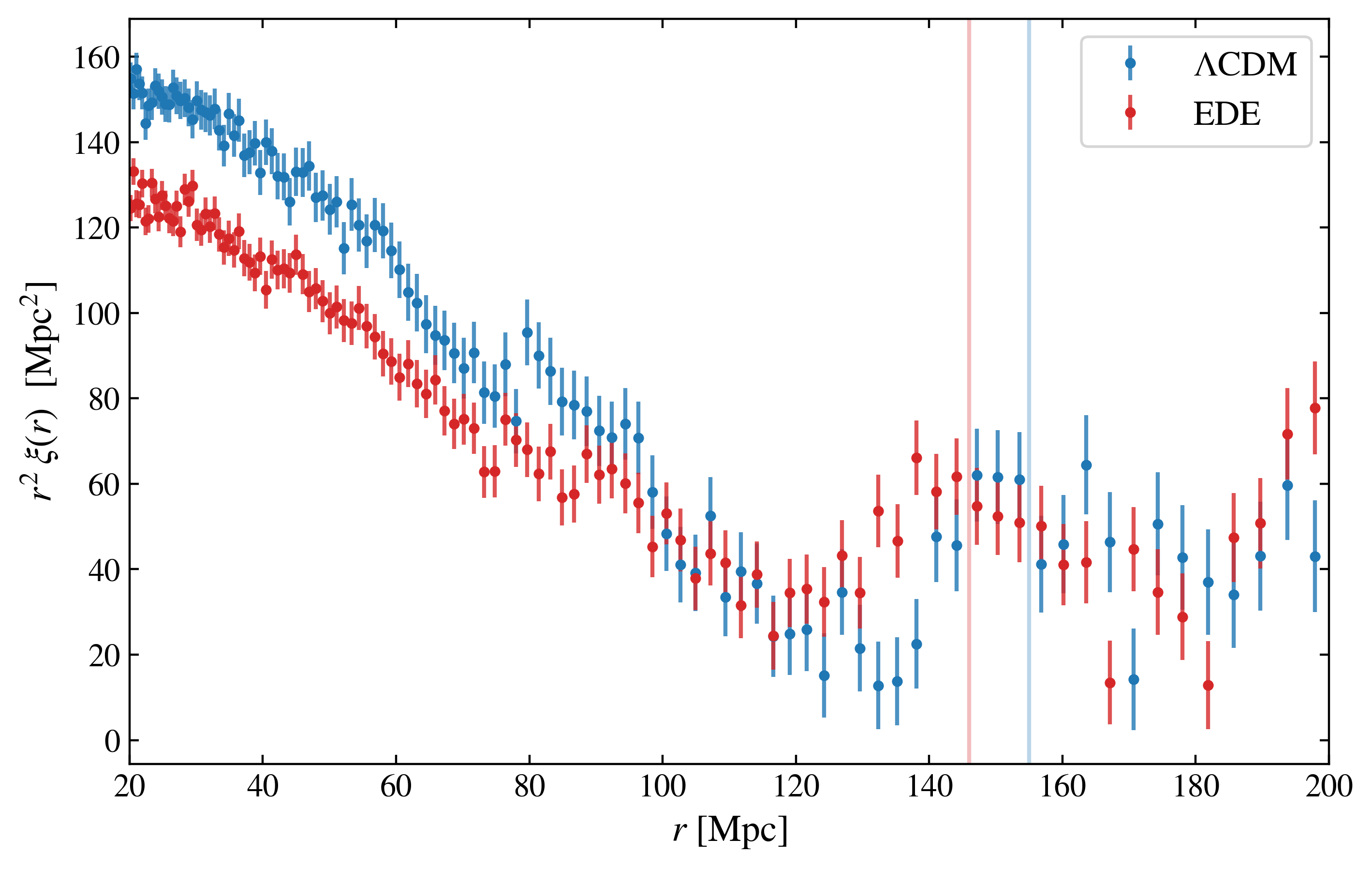}
    \caption{Two-point galaxy-galaxy correlation function, scaled by $r^2$. $\Lambda$CDM is shown in blue, while EDE is shown in red. Error bars are provided by TreeCorr. The vertical lines denote the rough position of the BAO peaks for $\Lambda$CDM (blue) and EDE (red).}
    \label{fig:gg-corr}
\end{figure}

Figure~\ref{fig:gg-corr} shows the correlation function scaled by $r^2$ to highlight differences between the models and emphasize the BAO peak. The correlation function for the $\Lambda$CDM model is overall higher than EDE, until large distances where the two are very close. There is a clear BAO peak for both simulations (marked in the figure) at a distance of around 145-155 Mpc. This is consistent with findings of the dark matter correlation function from \citet{Klypin_2020}, where the BAO peak was found at around 147 Mpc for both $\Lambda$CDM and EDE simulations. Similar to their findings, the BAO peak of EDE is shifted slightly from $\Lambda$CDM, yet here we find a shift to smaller radii rather than larger. It is unclear whether this shift is significant, or due to systematics. This is also in agreement with \citet{Fedeli_2009}, where it was found that $\Lambda$CDM produces higher correlations than EDE at all scales, but particularly on smaller scales.

\begin{figure}
    \centering
    \includegraphics[width=1\linewidth]{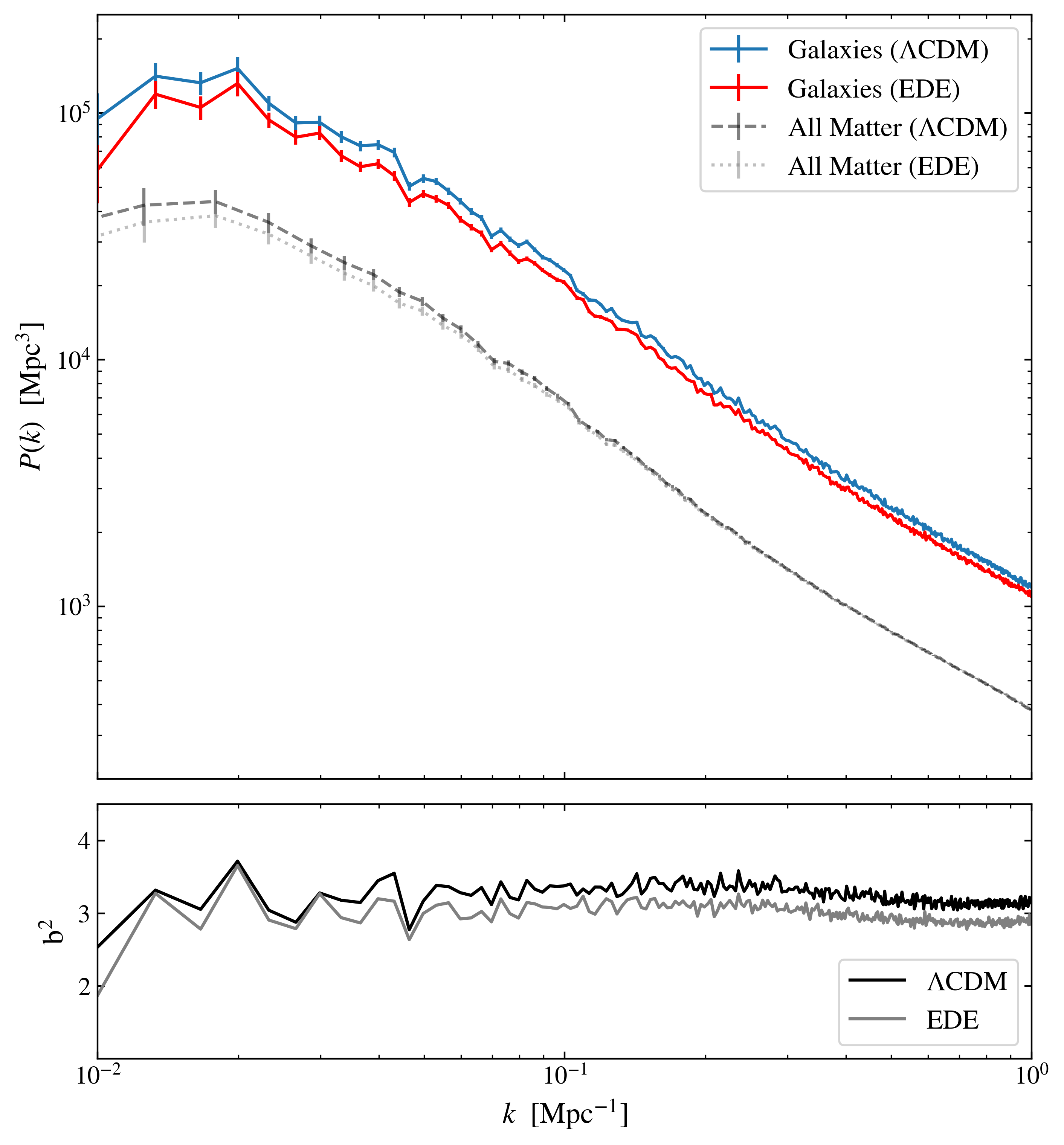}
    \caption{(Upper) Galaxy power spectra and matter power spectra. The galaxy power spectra are in color, with EDE in red and $\Lambda$CDM in blue. The matter power spectra for EDE and $\Lambda$CDM are shown by grey dotted and black dashed lines, respectively. (Lower) The linear bias squared for EDE in grey and $\Lambda$CDM in black.}
    \label{fig:gal_pow}
\end{figure}

Finally, we compute the galaxy power spectra by applying the Fourier transform directly on the mass field produced using a CiC mass deposition (see Section \ref{analysis-tools}). Figure~\ref{fig:gal_pow} shows the galaxy power spectra and the linear galaxy bias squared, defined as 
\begin{equation}
P_g(k) = b^2P_m(k).
\end{equation}
Overall the galaxy power spectra are higher than the matter power spectra, for both simulations. Once again we see the EDE power spectrum has a lower value than the $\Lambda$CDM model, however the squared biases are quite similar, with both hovering around 3 (or b $\sim 1.7$) for most $k$ values. \citet{Fontanot_2012} generate synthetic galaxy catalogs by applying a semi-analytic model to the results from N-body simulation. Their EDE implementation relies on the idea that a time-varying dark energy equation of state can mimic an early dark energy component. Generally, they find that the galaxy bias is higher for $\Lambda$CDM than for EDE, with increasing differences with higher redshifts, which corroborates our findings. Their biases are quite similar to ours, with small changes likely coming from the difference in the implementation and type of EDE model studied. Furthermore, our results are similar to those found for the LRG linear bias in DESI measurements \citep{Adame_2025}, and are near the values predicted for our $\sigma_8$ values from the 6dF Galaxy Survey \citep{Beutler_2012}.

\section{Data Release}
\label{sec:data}

As part of this publication we release several subsets of the simulation data. We release snapshots for $z\in \{0, 0.1, 0.5, 1.0\}$ for public use on the web-based portal introduced in \citet{Heitmann_2019}\footnote{https://cosmology.alcf.anl.gov}. To obtain the data, users must authenticate with a Globus identity, after which they can be downloaded using Globus transfer capabilities. 

In addition to full snapshots we make the data available on the OpenCosmo Compute Portal introduced in \citet{wells2026_opencosmo}, which allows users to select data relevant to their science and avoid downloading full snapshots. The access to the OpenCosmo Compute Portal is currently restricted to users with accounts at NERSC, the Argonne Leadership Computing Facility, or a DOE National Laboratory.

Available data products are halo catalogs, measured halo profiles, and particle data for high-mass halos. For details of these data products, see Sections 3.6.1-3.6.3 in \citet{wells2026_opencosmo}. Data retrieved from the OpenCosmo Compute Portal is returned in the OpenCosmo format and can readily be read and further analyzed with the OpenCosmo Python Toolkit\footnote{https://github.com/ArgonneCPAC/OpenCosmo}. Extensive examples of using the Toolkit, and comprehensive technical documentation, can be found at \url{https://argonnecpac.github.io/opencosmo-examples/}.

\section{Conclusions and Outlook}
\label{sec:conclusions}
In this paper we analyzed the two N-body simulations within the EDENS suite to explore a possible solution of the Hubble tension in the form of EDE. Our results show significant differences between an EDE and $\Lambda$CDM cosmology. However, as discussed in Sec.~\ref{sec:results}, many of these differences arise from the specific cosmology parameters, in particular $\sigma_8 \: \text{and} \: n_s$. 

We found that our EDE cosmology produces more massive halos that contain more galaxies than the $\Lambda$CDM cosmology, a significant difference near the BAO scale in the power spectra, and a clear difference in the galaxy correlation functions and power spectra. Our findings are in agreement with the literature, as \citet{Klypin_2020} found a difference of 10\% in the HMFs in the high mass regime and around 5\% in the nonlinear matter power spectra for $k\sim 0.1 \:\text{Mpc}^{-1}$ for $z=0$. The differences we find for the power spectrum measurements are slightly higher than their values at low wavenumbers, but lower at high wavenumbers. These differences are likely due to the small differences in cosmological parameters or simulation frameworks. We found the inverse relation to their shift in the BAO peak: the peak is shifted towards smaller distances in the correlation function, which corresponds to larger wavenumbers in Fourier space. We also found hints that the BAO peak may differ with redshift. While the concentration-mass relation is too close in the two models to be an accurate tracer of the cosmology, the HMFs, power spectra, and correlation functions deviate enough to possibly contextualize future observational data.

Observations of medium to high redshift galaxies could provide important constrains to distinguish the different cosmological models studied here, as several metric differences increase with redshift. While the correlation function was only calculated for $z=0.5$, we expect to see similar changes with redshift as in the HMFs. \citet{Halo_Galaxy_connect} and other papers have shown that there is a strong correlation between a galaxy's properties and its host halo's properties, making indirect observational measurements of halo mass possible. This will lead to improvements of the measurement of the HMFs from future observations with which the range of possible cosmological models can be further narrowed down. Furthermore, measurements of the 2-point correlation function of galaxies are already possible, and while we have not compared our simulations to these observations in this work, the significant differences at small $r$ and shift in the BAO peak are promising for determining which model may match observations better.

Some measurements we carried out led to very similar results for both models investigated, such as the concentration-mass relation and power spectra at large wavenumbers (small scales). The similar values show that while these metrics may not be powerful tracers to distinguish the models, the two models are similar enough that the EDE framework has promise as a solution to the Hubble Tension without losing the strengths of the $\Lambda$CDM cosmology. While more research is needed into other metrics and comparison to observations, this work suggests the likelihood that an EDE model would fit current observations while resolving the Hubble Tension and be detectable by future surveys. 

\begin{acknowledgments}
Work at Argonne National Laboratory was supported by the U.S. Department of Energy, Office of High Energy Physics. Argonne, a U.S. Department of Energy Office of Science Laboratory, is operated by UChicago Argonne LLC under contract no. DE-AC02-06CH11357. 

SH would like to thank Andrew Hearin for his advice on creating new models in HaloTools and suggestions on which models to implement. SH and PW would like to thank Patricia Larsen for her help with the galaxy power spectrum analysis. The authors would like to thank the HACC team for making the code readily available on Perlmutter.
\end{acknowledgments}

\bibliographystyle{aasjournal}
\bibliography{bibliography}

@ARTICLE{2026arXiv260318131W,
       author = {{Weiner}, Zachary J.},
        title = "{High-redshift physics from the acoustic scale}",
      journal = {arXiv e-prints},
         year = 2026,
        month = mar,
          eid = {arXiv:2603.18131},
        pages = {arXiv:2603.18131},
          doi = {10.48550/arXiv.2603.18131},
archivePrefix = {arXiv},
       eprint = {2603.18131},
 primaryClass = {astro-ph.CO},
       adsurl = {https://ui.adsabs.harvard.edu/abs/2026arXiv260318131W}
}

@ARTICLE{2025PhRvD.112h3511L,
       author = {{Lodha}, K. and {Calderon}, R. and {Matthewson}, W.~L. and {Shafieloo}, A. and {Ishak}, M. and {Pan}, J. and {Garcia-Quintero}, C. and {Huterer}, D. and {Valogiannis}, G. and {Ure{\~n}a-L{\'o}pez}, L.~A. and {Kamble}, N.~V. and {Parkinson}, D. and {Kim}, A.~G. and {Zhao}, G.~B. and {Cervantes-Cota}, J.~L. and {Rohlf}, J. and {Lozano-Rodr{\'\i}guez}, F. and {Rom{\'a}n-Herrera}, J.~O. and {Abdul-Karim}, M. and {Aguilar}, J. and {Ahlen}, S. and {Alves}, O. and {Andrade}, U. and {Armengaud}, E. and {Aviles}, A. and {Behera}, J. and {BenZvi}, S. and {Bianchi}, D. and {Brodzeller}, A. and {Brooks}, D. and {Burtin}, E. and {Canning}, R. and {Rosell}, A. Carnero and {Casas}, L. and {Castander}, F.~J. and {Charles}, M. and {Chaussidon}, E. and {Chaves-Montero}, J. and {Chebat}, D. and {Claybaugh}, T. and {Cole}, S. and {Cuceu}, A. and {Dawson}, K.~S. and {de la Macorra}, A. and {de Mattia}, A. and {Deiosso}, N. and {Demina}, R. and {Dey}, Arjun and {Dey}, Biprateep and {Ding}, Z. and {Doel}, P. and {Eisenstein}, D.~J. and {Elbers}, W. and {Ferraro}, S. and {Font-Ribera}, A. and {Forero-Romero}, J.~E. and {Garrison}, Lehman H. and {Gazta{\~n}aga}, E. and {Gil-Mar{\'\i}n}, H. and {Gontcho}, S. Gontcho A. and {Gonzalez-Morales}, A.~X. and {Gutierrez}, G. and {Guy}, J. and {Hahn}, C. and {Herbold}, M. and {Herrera-Alcantar}, H.~K. and {Honscheid}, K. and {Howlett}, C. and {Juneau}, S. and {Kehoe}, R. and {Kirkby}, D. and {Kisner}, T. and {Kremin}, A. and {Lahav}, O. and {Lamman}, C. and {Landriau}, M. and {Le Guillou}, L. and {Leauthaud}, A. and {Levi}, M.~E. and {Li}, Q. and {Magneville}, C. and {Manera}, M. and {Martini}, P. and {Meisner}, A. and {Mena-Fern{\'a}ndez}, J. and {Miquel}, R. and {Moustakas}, J. and {Santos}, D. Mu{\~n}oz and {Mu{\~n}oz-Guti{\'e}rrez}, A. and {Myers}, A.~D. and {Nadathur}, S. and {Niz}, G. and {Noriega}, H.~E. and {Paillas}, E. and {Palanque-Delabrouille}, N. and {Percival}, W.~J. and {Pieri}, Matthew M. and {Poppett}, C. and {Prada}, F. and {P{\'e}rez-Fern{\'a}ndez}, A. and {P{\'e}rez-R{\`a}fols}, I. and {Ram{\'\i}rez-P{\'e}rez}, C. and {Rashkovetskyi}, M. and {Ravoux}, C. and {Ross}, A.~J. and {Rossi}, G. and {Ruhlmann-Kleider}, V. and {Samushia}, L. and {Sanchez}, E. and {Schlegel}, D. and {Schubnell}, M. and {Seo}, H. and {Sinigaglia}, F. and {Sprayberry}, D. and {Tan}, T. and {Tarl{\'e}}, G. and {Taylor}, P. and {Turner}, W. and {Vargas-Maga{\~n}a}, M. and {Walther}, M. and {Weaver}, B.~A. and {Wolfson}, M. and {Y{\`e}che}, C. and {Zarrouk}, P. and {Zhou}, R. and {Zou}, H. and {DESI Collaboration}},
        title = "{Extended dark energy analysis using DESI DR2 BAO measurements}",
      journal = {\prd},
         year = 2025,
        month = oct,
       volume = {112},
       number = {8},
          eid = {083511},
        pages = {083511},
          doi = {10.1103/w4c6-1r5j},
archivePrefix = {arXiv},
       eprint = {2503.14743},
 primaryClass = {astro-ph.CO},
       adsurl = {https://ui.adsabs.harvard.edu/abs/2025PhRvD.112h3511L}
}

@ARTICLE{2025PhRvD.112h3513E,
       author = {{Elbers}, W. and {Aviles}, A. and {Noriega}, H.~E. and {Chebat}, D. and {Menegas}, A. and {Frenk}, C.~S. and {Garcia-Quintero}, C. and {Gonzalez}, D. and {Ishak}, M. and {Lahav}, O. and {Naidoo}, K. and {Niz}, G. and {Y{\`e}che}, C. and {Abdul-Karim}, M. and {Ahlen}, S. and {Alves}, O. and {Andrade}, U. and {Armengaud}, E. and {Behera}, J. and {BenZvi}, S. and {Bianchi}, D. and {Brieden}, S. and {Brodzeller}, A. and {Brooks}, D. and {Burtin}, E. and {Calderon}, R. and {Canning}, R. and {Carnero Rosell}, A. and {Casas}, L. and {Castander}, F.~J. and {Charles}, M. and {Chaussidon}, E. and {Chaves-Montero}, J. and {Claybaugh}, T. and {Cole}, S. and {Cooper}, A.~P. and {Cuceu}, A. and {Dawson}, K.~S. and {de la Macorra}, A. and {de Mattia}, A. and {Deiosso}, N. and {Dey}, A. and {Dey}, B. and {Ding}, Z. and {Doel}, P. and {Eisenstein}, D.~J. and {Ferraro}, S. and {Font-Ribera}, A. and {Forero-Romero}, J.~E. and {Garrison}, L.~H. and {Gazta{\~n}aga}, E. and {Gil-Mar{\'\i}n}, H. and {Gontcho}, S. Gontcho A. and {Gonzalez-Morales}, A.~X. and {Gutierrez}, G. and {He}, S. and {Herbold}, M. and {Herrera-Alcantar}, H.~K. and {Howlett}, C. and {Huterer}, D. and {Juneau}, S. and {Kehoe}, R. and {Kirkby}, D. and {Kisner}, T. and {Kremin}, A. and {Lamman}, C. and {Landriau}, M. and {Le Guillou}, L. and {Leauthaud}, A. and {Levi}, M.~E. and {Li}, Q. and {Lodha}, K. and {Magneville}, C. and {Manera}, M. and {Martini}, P. and {Matthewson}, W.~L. and {Meisner}, A. and {Mena-Fern{\'a}ndez}, J. and {Miquel}, R. and {Moustakas}, J. and {Nadathur}, S. and {Newman}, J.~A. and {Paillas}, E. and {Palanque-Delabrouille}, N. and {Percival}, W.~J. and {Pieri}, M.~M. and {Poppett}, C. and {Prada}, F. and {P{\'e}rez-R{\`a}fols}, I. and {Rabinowitz}, D. and {Ram{\'\i}rez-P{\'e}rez}, C. and {Rashkovetskyi}, M. and {Ravoux}, C. and {Rivera-Morales}, H. and {Rohlf}, J. and {Ross}, A.~J. and {Rossi}, G. and {Ruhlmann-Kleider}, V. and {Samushia}, L. and {Sanchez}, E. and {Schlegel}, D. and {Schubnell}, M. and {Seo}, H. and {Sinigaglia}, F. and {Sprayberry}, D. and {Tan}, T. and {Tarl{\'e}}, G. and {Taylor}, P. and {Turner}, W. and {Vargas-Maga{\~n}a}, M. and {Verde}, L. and {Walther}, M. and {Weaver}, B.~A. and {Whitford}, A. and {Wolfson}, M. and {Zarrouk}, P. and {Zhao}, C. and {Zhou}, R. and {Zou}, H. and {DESI Collaboration}},
        title = "{Constraints on neutrino physics from DESI DR2 BAO and DR1 full shape}",
      journal = {\prd},
         year = 2025,
        month = oct,
       volume = {112},
       number = {8},
          eid = {083513},
        pages = {083513},
          doi = {10.1103/w9pk-xsk7},
archivePrefix = {arXiv},
       eprint = {2503.14744},
 primaryClass = {astro-ph.CO},
       adsurl = {https://ui.adsabs.harvard.edu/abs/2025PhRvD.112h3513E}
}

@ARTICLE{2025PhRvD.112h3515A,
       author = {{Abdul Karim}, M. and {Aguilar}, J. and {Ahlen}, S. and {Alam}, S. and {Allen}, L. and {Allende Prieto}, C. and {Alves}, O. and {Anand}, A. and {Andrade}, U. and {Armengaud}, E. and {Aviles}, A. and {Bailey}, S. and {Baltay}, C. and {Bansal}, P. and {Bault}, A. and {Behera}, J. and {BenZvi}, S. and {Bianchi}, D. and {Blake}, C. and {Brieden}, S. and {Brodzeller}, A. and {Brooks}, D. and {Buckley-Geer}, E. and {Burtin}, E. and {Calderon}, R. and {Canning}, R. and {Rosell}, A. Carnero and {Carrilho}, P. and {Casas}, L. and {Castander}, F.~J. and {Charles}, M. and {Chaussidon}, E. and {Chaves-Montero}, J. and {Chebat}, D. and {Chen}, X. and {Claybaugh}, T. and {Cole}, S. and {Cooper}, A.~P. and {Cuceu}, A. and {Dawson}, K.~S. and {de la Macorra}, A. and {de Mattia}, A. and {Deiosso}, N. and {Della Costa}, J. and {Demina}, R. and {Dey}, A. and {Dey}, B. and {Ding}, Z. and {Doel}, P. and {Edelstein}, J. and {Eisenstein}, D.~J. and {Elbers}, W. and {Fagrelius}, P. and {Fanning}, K. and {Fern{\'a}ndez-Garc{\'\i}a}, E. and {Ferraro}, S. and {Font-Ribera}, A. and {Forero-Romero}, J.~E. and {Frenk}, C.~S. and {Garcia-Quintero}, C. and {Garrison}, L.~H. and {Gazta{\~n}aga}, E. and {Gil-Mar{\'\i}n}, H. and {Gontcho A Gontcho}, S. and {Gonzalez}, D. and {Gonzalez-Morales}, A.~X. and {Gordon}, C. and {Green}, D. and {Gutierrez}, G. and {Guy}, J. and {Hadzhiyska}, B. and {Hahn}, C. and {He}, S. and {Herbold}, M. and {Herrera-Alcantar}, H.~K. and {Ho}, M.-F. and {Honscheid}, K. and {Howlett}, C. and {Huterer}, D. and {Ishak}, M. and {Juneau}, S. and {Kamble}, N.~V. and {Kara{\c{c}}ayl{\i}}, N.~G. and {Kehoe}, R. and {Kent}, S. and {Kim}, A.~G. and {Kirkby}, D. and {Kisner}, T. and {Koposov}, S.~E. and {Kremin}, A. and {Krolewski}, A. and {Lahav}, O. and {Lamman}, C. and {Landriau}, M. and {Lang}, D. and {Lasker}, J. and {Le Goff}, J.~M. and {Le Guillou}, L. and {Leauthaud}, A. and {Levi}, M.~E. and {Li}, Q. and {Li}, T.~S. and {Lodha}, K. and {Lokken}, M. and {Lozano-Rodr{\'\i}guez}, F. and {Magneville}, C. and {Manera}, M. and {Martini}, P. and {Matthewson}, W.~L. and {Meisner}, A. and {Mena-Fern{\'a}ndez}, J. and {Menegas}, A. and {Mergulh{\~a}o}, T. and {Miquel}, R. and {Moustakas}, J. and {Mu{\~n}oz-Guti{\'e}rrez}, A. and {Mu{\~n}oz-Santos}, D. and {Myers}, A.~D. and {Nadathur}, S. and {Naidoo}, K. and {Napolitano}, L. and {Newman}, J.~A. and {Niz}, G. and {Noriega}, H.~E. and {Paillas}, E. and {Palanque-Delabrouille}, N. and {Pan}, J. and {Peacock}, J.~A. and {Pellejero Ibanez}, M. and {Percival}, W.~J. and {P{\'e}rez-Fern{\'a}ndez}, A. and {P{\'e}rez-R{\`a}fols}, I. and {Pieri}, M.~M. and {Poppett}, C. and {Prada}, F. and {Rabinowitz}, D. and {Raichoor}, A. and {Ram{\'\i}rez-P{\'e}rez}, C. and {Rashkovetskyi}, M. and {Ravoux}, C. and {Rich}, J. and {Rocher}, A. and {Rockosi}, C. and {Rohlf}, J. and {Rom{\'a}n-Herrera}, J.~O. and {Ross}, A.~J. and {Rossi}, G. and {Ruggeri}, R. and {Ruhlmann-Kleider}, V. and {Samushia}, L. and {Sanchez}, E. and {Sanders}, N. and {Schlegel}, D. and {Schubnell}, M. and {Seo}, H. and {Shafieloo}, A. and {Sharples}, R. and {Silber}, J. and {Sinigaglia}, F. and {Sprayberry}, D. and {Tan}, T. and {Tarl{\'e}}, G. and {Taylor}, P. and {Turner}, W. and {Ure{\~n}a-L{\'o}pez}, L.~A. and {Vaisakh}, R. and {Valdes}, F. and {Valogiannis}, G. and {Vargas-Maga{\~n}a}, M. and {Verde}, L. and {Walther}, M. and {Weaver}, B.~A. and {Weinberg}, D.~H. and {White}, M. and {Wolfson}, M. and {Y{\`e}che}, C. and {Yu}, J. and {Zaborowski}, E.~A. and {Zarrouk}, P. and {Zhai}, Z. and {Zhang}, H. and {Zhao}, C. and {Zhao}, G.~B. and {Zhou}, R. and {Zou}, H. and {DESI Collaboration}},
        title = "{DESI DR2 results. II. Measurements of baryon acoustic oscillations and cosmological constraints}",
      journal = {\prd},
         year = 2025,
        month = oct,
       volume = {112},
       number = {8},
          eid = {083515},
        pages = {083515},
          doi = {10.1103/tr6y-kpc6},
archivePrefix = {arXiv},
       eprint = {2503.14738},
 primaryClass = {astro-ph.CO},
       adsurl = {https://ui.adsabs.harvard.edu/abs/2025PhRvD.112h3515A}
}

@ARTICLE{2016A&A...594A..11P,
       author = {{Planck Collaboration} and {Aghanim}, N. and {Arnaud}, M. and {Ashdown}, M. and {Aumont}, J. and {Baccigalupi}, C. and {Banday}, A.~J. and {Barreiro}, R.~B. and {Bartlett}, J.~G. and {Bartolo}, N. and {Battaner}, E. and {Benabed}, K. and {Beno{\^\i}t}, A. and {Benoit-L{\'e}vy}, A. and {Bernard}, J.-P. and {Bersanelli}, M. and {Bielewicz}, P. and {Bock}, J.~J. and {Bonaldi}, A. and {Bonavera}, L. and {Bond}, J.~R. and {Borrill}, J. and {Bouchet}, F.~R. and {Boulanger}, F. and {Bucher}, M. and {Burigana}, C. and {Butler}, R.~C. and {Calabrese}, E. and {Cardoso}, J.-F. and {Catalano}, A. and {Challinor}, A. and {Chiang}, H.~C. and {Christensen}, P.~R. and {Clements}, D.~L. and {Colombo}, L.~P.~L. and {Combet}, C. and {Coulais}, A. and {Crill}, B.~P. and {Curto}, A. and {Cuttaia}, F. and {Danese}, L. and {Davies}, R.~D. and {Davis}, R.~J. and {de Bernardis}, P. and {de Rosa}, A. and {de Zotti}, G. and {Delabrouille}, J. and {D{\'e}sert}, F.-X. and {Di Valentino}, E. and {Dickinson}, C. and {Diego}, J.~M. and {Dolag}, K. and {Dole}, H. and {Donzelli}, S. and {Dor{\'e}}, O. and {Douspis}, M. and {Ducout}, A. and {Dunkley}, J. and {Dupac}, X. and {Efstathiou}, G. and {Elsner}, F. and {En{\ss}lin}, T.~A. and {Eriksen}, H.~K. and {Fergusson}, J. and {Finelli}, F. and {Forni}, O. and {Frailis}, M. and {Fraisse}, A.~A. and {Franceschi}, E. and {Frejsel}, A. and {Galeotta}, S. and {Galli}, S. and {Ganga}, K. and {Gauthier}, C. and {Gerbino}, M. and {Giard}, M. and {Gjerl{\o}w}, E. and {Gonz{\'a}lez-Nuevo}, J. and {G{\'o}rski}, K.~M. and {Gratton}, S. and {Gregorio}, A. and {Gruppuso}, A. and {Gudmundsson}, J.~E. and {Hamann}, J. and {Hansen}, F.~K. and {Harrison}, D.~L. and {Helou}, G. and {Henrot-Versill{\'e}}, S. and {Hern{\'a}ndez-Monteagudo}, C. and {Herranz}, D. and {Hildebrandt}, S.~R. and {Hivon}, E. and {Holmes}, W.~A. and {Hornstrup}, A. and {Huffenberger}, K.~M. and {Hurier}, G. and {Jaffe}, A.~H. and {Jones}, W.~C. and {Juvela}, M. and {Keih{\"a}nen}, E. and {Keskitalo}, R. and {Kiiveri}, K. and {Knoche}, J. and {Knox}, L. and {Kunz}, M. and {Kurki-Suonio}, H. and {Lagache}, G. and {L{\"a}hteenm{\"a}ki}, A. and {Lamarre}, J.-M. and {Lasenby}, A. and {Lattanzi}, M. and {Lawrence}, C.~R. and {Le Jeune}, M. and {Leonardi}, R. and {Lesgourgues}, J. and {Levrier}, F. and {Lewis}, A. and {Liguori}, M. and {Lilje}, P.~B. and {Lilley}, M. and {Linden-V{\o}rnle}, M. and {Lindholm}, V. and {L{\'o}pez-Caniego}, M. and {Mac{\'\i}as-P{\'e}rez}, J.~F. and {Maffei}, B. and {Maggio}, G. and {Maino}, D. and {Mandolesi}, N. and {Mangilli}, A. and {Maris}, M. and {Martin}, P.~G. and {Mart{\'\i}nez-Gonz{\'a}lez}, E. and {Masi}, S. and {Matarrese}, S. and {Meinhold}, P.~R. and {Melchiorri}, A. and {Migliaccio}, M. and {Millea}, M. and {Mitra}, S. and {Miville-Desch{\^e}nes}, M.-A. and {Moneti}, A. and {Montier}, L. and {Morgante}, G. and {Mortlock}, D. and {Mottet}, S. and {Munshi}, D. and {Murphy}, J.~A. and {Narimani}, A. and {Naselsky}, P. and {Nati}, F. and {Natoli}, P. and {Noviello}, F. and {Novikov}, D. and {Novikov}, I. and {Oxborrow}, C.~A. and {Paci}, F. and {Pagano}, L. and {Pajot}, F. and {Paoletti}, D. and {Partridge}, B. and {Pasian}, F. and {Patanchon}, G. and {Pearson}, T.~J. and {Perdereau}, O. and {Perotto}, L. and {Pettorino}, V. and {Piacentini}, F. and {Piat}, M. and {Pierpaoli}, E. and {Pietrobon}, D. and {Plaszczynski}, S. and {Pointecouteau}, E. and {Polenta}, G. and {Ponthieu}, N. and {Pratt}, G.~W. and {Prunet}, S. and {Puget}, J.-L. and {Rachen}, J.~P. and {Reinecke}, M. and {Remazeilles}, M. and {Renault}, C. and {Renzi}, A. and {Ristorcelli}, I. and {Rocha}, G. and {Rossetti}, M. and {Roudier}, G. and {Rouill{\'e} d'Orfeuil}, B. and {Rubi{\~n}o-Mart{\'\i}n}, J.~A. and {Rusholme}, B. and {Salvati}, L. and {Sandri}, M. and {Santos}, D. and {Savelainen}, M. and {Savini}, G. and {Scott}, D. and {Serra}, P. and {Spencer}, L.~D. and {Spinelli}, M. and {Stolyarov}, V. and {Stompor}, R.},
        title = "{Planck 2015 results. XI. CMB power spectra, likelihoods, and robustness of parameters}",
      journal = {\aap},
         year = 2016,
        month = sep,
       volume = {594},
          eid = {A11},
        pages = {A11},
          doi = {10.1051/0004-6361/201526926},
archivePrefix = {arXiv},
       eprint = {1507.02704},
 primaryClass = {astro-ph.CO},
       adsurl = {https://ui.adsabs.harvard.edu/abs/2016A&A...594A..11P}
}

@ARTICLE{2019ApJ...876...85R,
       author = {{Riess}, Adam G. and {Casertano}, Stefano and {Yuan}, Wenlong and {Macri}, Lucas M. and {Scolnic}, Dan},
        title = "{Large Magellanic Cloud Cepheid Standards Provide a 1\% Foundation for the Determination of the Hubble Constant and Stronger Evidence for Physics beyond {\ensuremath{\Lambda}}CDM}",
      journal = {\apj},
         year = 2019,
        month = may,
       volume = {876},
       number = {1},
          eid = {85},
        pages = {85},
          doi = {10.3847/1538-4357/ab1422},
archivePrefix = {arXiv},
       eprint = {1903.07603},
 primaryClass = {astro-ph.CO},
       adsurl = {https://ui.adsabs.harvard.edu/abs/2019ApJ...876...85R}
}

@ARTICLE{2011MNRAS.416.3017B,
       author = {{Beutler}, Florian and {Blake}, Chris and {Colless}, Matthew and {Jones}, D. Heath and {Staveley-Smith}, Lister and {Campbell}, Lachlan and {Parker}, Quentin and {Saunders}, Will and {Watson}, Fred},
        title = "{The 6dF Galaxy Survey: baryon acoustic oscillations and the local Hubble constant}",
      journal = {\mnras},
         year = 2011,
        month = oct,
       volume = {416},
       number = {4},
        pages = {3017-3032},
          doi = {10.1111/j.1365-2966.2011.19250.x},
archivePrefix = {arXiv},
       eprint = {1106.3366},
 primaryClass = {astro-ph.CO},
       adsurl = {https://ui.adsabs.harvard.edu/abs/2011MNRAS.416.3017B}
}

@ARTICLE{2015MNRAS.449..835R,
       author = {{Ross}, Ashley J. and {Samushia}, Lado and {Howlett}, Cullan and {Percival}, Will J. and {Burden}, Angela and {Manera}, Marc},
        title = "{The clustering of the SDSS DR7 main Galaxy sample - I. A 4 per cent distance measure at z = 0.15}",
      journal = {\mnras},
         year = 2015,
        month = may,
       volume = {449},
       number = {1},
        pages = {835-847},
          doi = {10.1093/mnras/stv154},
archivePrefix = {arXiv},
       eprint = {1409.3242},
 primaryClass = {astro-ph.CO},
       adsurl = {https://ui.adsabs.harvard.edu/abs/2015MNRAS.449..835R}
}

@ARTICLE{2017MNRAS.470.2617A,
       author = {{Alam}, Shadab and {Ata}, Metin and {Bailey}, Stephen and {Beutler}, Florian and {Bizyaev}, Dmitry and {Blazek}, Jonathan A. and {Bolton}, Adam S. and {Brownstein}, Joel R. and {Burden}, Angela and {Chuang}, Chia-Hsun and {Comparat}, Johan and {Cuesta}, Antonio J. and {Dawson}, Kyle S. and {Eisenstein}, Daniel J. and {Escoffier}, Stephanie and {Gil-Mar{\'\i}n}, H{\'e}ctor and {Grieb}, Jan Niklas and {Hand}, Nick and {Ho}, Shirley and {Kinemuchi}, Karen and {Kirkby}, David and {Kitaura}, Francisco and {Malanushenko}, Elena and {Malanushenko}, Viktor and {Maraston}, Claudia and {McBride}, Cameron K. and {Nichol}, Robert C. and {Olmstead}, Matthew D. and {Oravetz}, Daniel and {Padmanabhan}, Nikhil and {Palanque-Delabrouille}, Nathalie and {Pan}, Kaike and {Pellejero-Ibanez}, Marcos and {Percival}, Will J. and {Petitjean}, Patrick and {Prada}, Francisco and {Price-Whelan}, Adrian M. and {Reid}, Beth A. and {Rodr{\'\i}guez-Torres}, Sergio A. and {Roe}, Natalie A. and {Ross}, Ashley J. and {Ross}, Nicholas P. and {Rossi}, Graziano and {Rubi{\~n}o-Mart{\'\i}n}, Jose Alberto and {Saito}, Shun and {Salazar-Albornoz}, Salvador and {Samushia}, Lado and {S{\'a}nchez}, Ariel G. and {Satpathy}, Siddharth and {Schlegel}, David J. and {Schneider}, Donald P. and {Sc{\'o}ccola}, Claudia G. and {Seo}, Hee-Jong and {Sheldon}, Erin S. and {Simmons}, Audrey and {Slosar}, An{\v{z}}e and {Strauss}, Michael A. and {Swanson}, Molly E.~C. and {Thomas}, Daniel and {Tinker}, Jeremy L. and {Tojeiro}, Rita and {Maga{\~n}a}, Mariana Vargas and {Vazquez}, Jose Alberto and {Verde}, Licia and {Wake}, David A. and {Wang}, Yuting and {Weinberg}, David H. and {White}, Martin and {Wood-Vasey}, W. Michael and {Y{\`e}che}, Christophe and {Zehavi}, Idit and {Zhai}, Zhongxu and {Zhao}, Gong-Bo},
        title = "{The clustering of galaxies in the completed SDSS-III Baryon Oscillation Spectroscopic Survey: cosmological analysis of the DR12 galaxy sample}",
      journal = {\mnras},
         year = 2017,
        month = sep,
       volume = {470},
       number = {3},
        pages = {2617-2652},
          doi = {10.1093/mnras/stx721},
archivePrefix = {arXiv},
       eprint = {1607.03155},
 primaryClass = {astro-ph.CO},
       adsurl = {https://ui.adsabs.harvard.edu/abs/2017MNRAS.470.2617A}
}

@ARTICLE{2018ApJ...859..101S,
       author = {{Scolnic}, D.~M. and {Jones}, D.~O. and {Rest}, A. and {Pan}, Y.~C. and {Chornock}, R. and {Foley}, R.~J. and {Huber}, M.~E. and {Kessler}, R. and {Narayan}, G. and {Riess}, A.~G. and {Rodney}, S. and {Berger}, E. and {Brout}, D.~J. and {Challis}, P.~J. and {Drout}, M. and {Finkbeiner}, D. and {Lunnan}, R. and {Kirshner}, R.~P. and {Sanders}, N.~E. and {Schlafly}, E. and {Smartt}, S. and {Stubbs}, C.~W. and {Tonry}, J. and {Wood-Vasey}, W.~M. and {Foley}, M. and {Hand}, J. and {Johnson}, E. and {Burgett}, W.~S. and {Chambers}, K.~C. and {Draper}, P.~W. and {Hodapp}, K.~W. and {Kaiser}, N. and {Kudritzki}, R.~P. and {Magnier}, E.~A. and {Metcalfe}, N. and {Bresolin}, F. and {Gall}, E. and {Kotak}, R. and {McCrum}, M. and {Smith}, K.~W.},
        title = "{The Complete Light-curve Sample of Spectroscopically Confirmed SNe Ia from Pan-STARRS1 and Cosmological Constraints from the Combined Pantheon Sample}",
      journal = {\apj},
         year = 2018,
        month = jun,
       volume = {859},
       number = {2},
          eid = {101},
        pages = {101},
          doi = {10.3847/1538-4357/aab9bb},
archivePrefix = {arXiv},
       eprint = {1710.00845},
 primaryClass = {astro-ph.CO},
       adsurl = {https://ui.adsabs.harvard.edu/abs/2018ApJ...859..101S}
}

@ARTICLE{2026arXiv260713282S,
       author = {{Sch{\"o}neberg}, Nils and {Poulin}, Vivian and {Ferrari}, Angelo G. and {Finelli}, Fabio and {Lesgourgues}, Julien and {Morelli}, Luca and {Mosbech}, Markus R. and {Sharma}, Ravi Kumar and {Simon}, Th{\'e}o},
        title = "{The $H_0$ World Cup. I. Summary of the baseline group stage results}",
      journal = {arXiv e-prints},
         year = 2026,
        month = jul,
          eid = {arXiv:2607.13282},
        pages = {arXiv:2607.13282},
          doi = {10.48550/arXiv.2607.13282},
archivePrefix = {arXiv},
       eprint = {2607.13282},
 primaryClass = {astro-ph.CO},
       adsurl = {https://ui.adsabs.harvard.edu/abs/2026arXiv260713282S}
}

@ARTICLE{2026arXiv260713283S,
       author = {{Sch{\"o}neberg}, Nils and {Poulin}, Vivian and {Ferrari}, Angelo G. and {Finelli}, Fabio and {Lesgourgues}, Julien and {Morelli}, Luca and {Mosbech}, Markus R. and {Sharma}, Ravi Kumar and {Simon}, Th{\'e}o},
        title = "{The $H_0$ world cup. II. A comprehensive competition between proposed Hubble tension solutions}",
      journal = {arXiv e-prints},
         year = 2026,
        month = jul,
          eid = {arXiv:2607.13283},
        pages = {arXiv:2607.13283},
          doi = {10.48550/arXiv.2607.13283},
archivePrefix = {arXiv},
       eprint = {2607.13283},
 primaryClass = {astro-ph.CO},
       adsurl = {https://ui.adsabs.harvard.edu/abs/2026arXiv260713283S}
}

@ARTICLE{2000MNRAS.318.1144P,
       author = {{Peacock}, J.~A. and {Smith}, R.~E.},
        title = "{Halo occupation numbers and galaxy bias}",
      journal = {\mnras},
         year = 2000,
        month = nov,
       volume = {318},
       number = {4},
        pages = {1144-1156},
          doi = {10.1046/j.1365-8711.2000.03779.x},
archivePrefix = {arXiv},
       eprint = {astro-ph/0005010},
 primaryClass = {astro-ph},
       adsurl = {https://ui.adsabs.harvard.edu/abs/2000MNRAS.318.1144P}
}

@ARTICLE{2002ApJ...575..587B,
       author = {{Berlind}, Andreas A. and {Weinberg}, David H.},
        title = "{The Halo Occupation Distribution: Toward an Empirical Determination of the Relation between Galaxies and Mass}",
      journal = {\apj},
         year = 2002,
        month = aug,
       volume = {575},
       number = {2},
        pages = {587-616},
          doi = {10.1086/341469},
archivePrefix = {arXiv},
       eprint = {astro-ph/0109001},
 primaryClass = {astro-ph},
       adsurl = {https://ui.adsabs.harvard.edu/abs/2002ApJ...575..587B}
}

@ARTICLE{2000MNRAS.318..203S,
       author = {{Seljak}, Uro{\v{s}}},
        title = "{Analytic model for galaxy and dark matter clustering}",
      journal = {\mnras},
         year = 2000,
        month = oct,
       volume = {318},
       number = {1},
        pages = {203-213},
          doi = {10.1046/j.1365-8711.2000.03715.x},
archivePrefix = {arXiv},
       eprint = {astro-ph/0001493},
 primaryClass = {astro-ph},
       adsurl = {https://ui.adsabs.harvard.edu/abs/2000MNRAS.318..203S}
}

@ARTICLE{2001ApJ...546...20S,
       author = {{Scoccimarro}, Rom{\'a}n and {Sheth}, Ravi K. and {Hui}, Lam and {Jain}, Bhuvnesh},
        title = "{How Many Galaxies Fit in a Halo? Constraints on Galaxy Formation Efficiency from Spatial Clustering}",
      journal = {\apj},
         year = 2001,
        month = jan,
       volume = {546},
       number = {1},
        pages = {20-34},
          doi = {10.1086/318261},
archivePrefix = {arXiv},
       eprint = {astro-ph/0006319},
 primaryClass = {astro-ph},
       adsurl = {https://ui.adsabs.harvard.edu/abs/2001ApJ...546...20S}
}

@ARTICLE{Yuan_2021,
   title={<scp>AbacusHOD</scp>: a highly efficient extended multitracer HOD framework and its application to BOSS and eBOSS data},
   volume={510},
   ISSN={1365-2966},
   url={http://dx.doi.org/10.1093/mnras/stab3355},
   DOI={10.1093/mnras/stab3355},
   number={3},
   journal={Monthly Notices of the Royal Astronomical Society},
   publisher={Oxford University Press (OUP)},
   author={Yuan, Sihan and Garrison, Lehman H and Hadzhiyska, Boryana and Bose, Sownak and Eisenstein, Daniel J},
   year={2021},
   month=nov, pages={3301–3320} }

@ARTICLE{White_2011,
doi = {10.1088/0004-637X/728/2/126},
url = {https://doi.org/10.1088/0004-637X/728/2/126},
year = {2011},
month = {jan},
publisher = {The American Astronomical Society},
volume = {728},
number = {2},
pages = {126},
author = {White, Martin and Blanton, M. and Bolton, A. and Schlegel, D. and Tinker, J. and Berlind, A. and da Costa, L. and Kazin, E. and Lin, Y.-T. and Maia, M. and McBride, C. K. and Padmanabhan, N. and Parejko, J. and Percival, W. and Prada, F. and Ramos, B. and Sheldon, E. and de Simoni, F. and Skibba, R. and Thomas, D. and Wake, D. and Zehavi, I. and Zheng, Z. and Nichol, R. and Schneider, Donald P. and Strauss, Michael A. and Weaver, B. A. and Weinberg, David H.},
title = {THE CLUSTERING OF MASSIVE GALAXIES AT z ∼ 0.5 FROM THE FIRST SEMESTER OF BOSS DATA},
journal = {The Astrophysical Journal}
}

@ARTICLE{Hearin_2017,
   title={Forward Modeling of Large-scale Structure: An Open-source Approach with Halotools},
   volume={154},
   ISSN={1538-3881},
   url={http://dx.doi.org/10.3847/1538-3881/aa859f},
   DOI={10.3847/1538-3881/aa859f},
   number={5},
   journal={The Astronomical Journal},
   publisher={American Astronomical Society},
   author={Hearin, Andrew P. and Campbell, Duncan and Tollerud, Erik and Behroozi, Peter and Diemer, Benedikt and Goldbaum, Nathan J. and Jennings, Elise and Leauthaud, Alexie and Mao, Yao-Yuan and More, Surhud and Parejko, John and Sinha, Manodeep and {Sipöcz}, Brigitta and Zentner, Andrew},
   year={2017},
   month=Oct, pages={190} }

@ARTICLE{Asgari_2023,
   title={The halo model for cosmology: a pedagogical review},
   volume={6},
   ISSN={2565-6120},
   url={http://dx.doi.org/10.21105/astro.2303.08752},
   DOI={10.21105/astro.2303.08752},
   journal={The Open Journal of Astrophysics},
   publisher={Maynooth University},
   author={Asgari, Marika and Mead, Alexander J. and Heymans, Catherine},
   year={2023},
   month=Nov }

@ARTICLE{Zhou_2023,
   title={Target Selection and Validation of DESI Luminous Red Galaxies},
   volume={165},
   ISSN={1538-3881},
   url={http://dx.doi.org/10.3847/1538-3881/aca5fb},
   DOI={10.3847/1538-3881/aca5fb},
   number={2},
   journal={The Astronomical Journal},
   publisher={American Astronomical Society},
   author={Zhou, Rongpu and Dey, Biprateep and Newman, Jeffrey A. and Eisenstein, Daniel J. and Dawson, K. and Bailey, S. and Berti, A. and Guy, J. and Lan, Ting-Wen and Zou, H. and Aguilar, J. and Ahlen, S. and Alam, Shadab and Brooks, D. and de la Macorra, A. and Dey, A. and Dhungana, G. and Fanning, K. and Font-Ribera, A. and Gontcho, S. Gontcho A. and Honscheid, K. and Ishak, Mustapha and Kisner, T. and {Kovács}, A. and Kremin, A. and Landriau, M. and Levi, Michael E. and Magneville, C. and Manera, Marc and Martini, P. and Meisner, Aaron M. and Miquel, R. and Moustakas, J. and Myers, Adam D. and Nie, Jundan and Palanque-Delabrouille, N. and Percival, W. J. and Poppett, C. and Prada, F. and Raichoor, A. and Ross, A. J. and Schlafly, E. and Schlegel, D. and Schubnell, M. and {Tarlé}, Gregory and Weaver, B. A. and Wechsler, R. H. and {Yéche}, Christophe and Zhou, Zhimin},
   year={2023},
   month=Jan, pages={58} }

@ARTICLE{2016NewA...42...49H,
       author = {{Habib}, Salman and {Pope}, Adrian and {Finkel}, Hal and {Frontiere}, Nicholas and {Heitmann}, Katrin and {Daniel}, David and {Fasel}, Patricia and {Morozov}, Vitali and {Zagaris}, George and {Peterka}, Tom and {Vishwanath}, Venkatram and {Luki{\'c}}, Zarija and {Sehrish}, Saba and {Liao}, Wei-keng},
        title = "{HACC: Simulating sky surveys on state-of-the-art supercomputing architectures}",
      journal = {\na},
         year = 2016,
        month = jan,
       volume = {42},
        pages = {49-65},
          doi = {10.1016/j.newast.2015.06.003},
archivePrefix = {arXiv},
       eprint = {1410.2805},
 primaryClass = {astro-ph.IM},
       adsurl = {https://ui.adsabs.harvard.edu/abs/2016NewA...42...49H}
}

@ARTICLE{2012arXiv1211.4864H,
       author = {{Habib}, Salman and {Morozov}, Vitali and {Finkel}, Hal and {Pope}, Adrian and {Heitmann}, Katrin and {Kumaran}, Kalyan and {Peterka}, Tom and {Insley}, Joe and {Daniel}, David and {Fasel}, Patricia and {Frontiere}, Nicholas and {Lukic}, Zarija},
        title = "{The Universe at Extreme Scale: Multi-Petaflop Sky Simulation on the BG/Q}",
      journal = {arXiv e-prints},
         year = 2012,
        month = nov,
          eid = {arXiv:1211.4864},
        pages = {arXiv:1211.4864},
          doi = {10.48550/arXiv.1211.4864},
archivePrefix = {arXiv},
       eprint = {1211.4864},
 primaryClass = {cs.DC},
       adsurl = {https://ui.adsabs.harvard.edu/abs/2012arXiv1211.4864H}
}

@ARTICLE{2019ApJS..245...16H,
       author = {{Heitmann}, Katrin and {Finkel}, Hal and {Pope}, Adrian and {Morozov}, Vitali and {Frontiere}, Nicholas and {Habib}, Salman and {Rangel}, Esteban and {Uram}, Thomas and {Korytov}, Danila and {Child}, Hillary and {Flender}, Samuel and {Insley}, Joe and {Rizzi}, Silvio},
        title = "{The Outer Rim Simulation: A Path to Many-core Supercomputers}",
      journal = {\apjs},
         year = 2019,
        month = nov,
       volume = {245},
       number = {1},
          eid = {16},
        pages = {16},
          doi = {10.3847/1538-4365/ab4da1},
archivePrefix = {arXiv},
       eprint = {1904.11970},
 primaryClass = {astro-ph.CO},
       adsurl = {https://ui.adsabs.harvard.edu/abs/2019ApJS..245...16H}
}

@ARTICLE{2023ApJS..264...34F,
       author = {{Frontiere}, Nicholas and {Emberson}, J.~D. and {Buehlmann}, Michael and {Adamo}, Joseph and {Habib}, Salman and {Heitmann}, Katrin and {Faucher-Gigu{\`e}re}, Claude-Andr{\'e}},
        title = "{Simulating Hydrodynamics in Cosmology with CRK-HACC}",
      journal = {\apjs},
         year = 2023,
        month = feb,
       volume = {264},
       number = {2},
          eid = {34},
        pages = {34},
          doi = {10.3847/1538-4365/aca58d},
archivePrefix = {arXiv},
       eprint = {2202.02840},
 primaryClass = {astro-ph.CO},
       adsurl = {https://ui.adsabs.harvard.edu/abs/2023ApJS..264...34F}
}

@INPROCEEDINGS{2025hpcn.conf...25F,
       author = {{Frontiere}, Nicholas and {Emberson}, J.~D. and {Buehlmann}, Michael and {Rangel}, Esteban M. and {Habib}, Salman and {Heitmann}, Katrin and {Larsen}, Patricia and {Morozov}, Vitali and {Pope}, Adrian and {Faucher-Gigu{\`e}re}, Claude-Andr{\'e} and {Georgiadou}, Antigoni and {Lebrun-Grandi{\'e}}, Damien and {Prokopenko}, Andrey},
        title = "{Cosmological Hydrodynamics at Exascale: A Trillion-Particle Leap in Capability}",
    booktitle = {SC '25: Proceedings of the International Conference for High Performance Computing},
         year = 2025,
        month = nov,
        pages = {25-35},
          doi = {10.1145/3712285.3771786},
archivePrefix = {arXiv},
       eprint = {2510.03557},
 primaryClass = {cs.DC},
       adsurl = {https://ui.adsabs.harvard.edu/abs/2025hpcn.conf...25F}
}

@ARTICLE{2011MNRAS.415.2293K,
       author = {{Knebe}, Alexander and {Knollmann}, Steffen R. and {Muldrew}, Stuart I. and {Pearce}, Frazer R. and {Aragon-Calvo}, Miguel Angel and {Ascasibar}, Yago and {Behroozi}, Peter S. and {Ceverino}, Daniel and {Colombi}, Stephane and {Diemand}, Juerg and {Dolag}, Klaus and {Falck}, Bridget L. and {Fasel}, Patricia and {Gardner}, Jeff and {Gottl{\"o}ber}, Stefan and {Hsu}, Chung-Hsing and {Iannuzzi}, Francesca and {Klypin}, Anatoly and {Luki{\'c}}, Zarija and {Maciejewski}, Michal and {McBride}, Cameron and {Neyrinck}, Mark C. and {Planelles}, Susana and {Potter}, Doug and {Quilis}, Vicent and {Rasera}, Yann and {Read}, Justin I. and {Ricker}, Paul M. and {Roy}, Fabrice and {Springel}, Volker and {Stadel}, Joachim and {Stinson}, Greg and {Sutter}, P.~M. and {Turchaninov}, Victor and {Tweed}, Dylan and {Yepes}, Gustavo and {Zemp}, Marcel},
        title = "{Haloes gone MAD: The Halo-Finder Comparison Project}",
      journal = {\mnras},
         year = 2011,
        month = aug,
       volume = {415},
       number = {3},
        pages = {2293-2318},
          doi = {10.1111/j.1365-2966.2011.18858.x},
archivePrefix = {arXiv},
       eprint = {1104.0949},
 primaryClass = {astro-ph.CO},
       adsurl = {https://ui.adsabs.harvard.edu/abs/2011MNRAS.415.2293K}
}

@ARTICLE{Heitmann_2021,
    doi = {10.3847/1538-4365/abcc67},
    url = {https://dx.doi.org/10.3847/1538-4365/abcc67},
    year = {2021},
    month = {jan},
    publisher = {The American Astronomical Society},
    volume = {252},
    number = {2},
    pages = {19},
    author = {Heitmann, Katrin and Frontiere, Nicholas and Rangel, Esteban and Larsen, Patricia and Pope, Adrian and Sultan, Imran and Uram, Thomas and Habib, Salman and Finkel, Hal and Korytov, Danila and Kovacs, Eve and Rizzi, Silvio and Insley, Joe and Knowles, Janet Y. K.},
    title = {The Last Journey. I. An Extreme-scale Simulation on the Mira Supercomputer},
    journal = {The Astrophysical Journal Supplement Series}}

@ARTICLE{Smith_2020,
   title={Oscillating scalar fields and the Hubble tension: A resolution with novel signatures},
   volume={101},
   ISSN={2470-0029},
   url={http://dx.doi.org/10.1103/PhysRevD.101.063523},
   DOI={10.1103/physrevd.101.063523},
   number={6},
   journal={Physical Review D},
   publisher={American Physical Society (APS)},
   author={Smith, Tristan L. and Poulin, Vivian and Amin, Mustafa A.},
   year={2020},
   month=mar }

@ARTICLE{Di_Valentino_2021,
   title={A combined analysis of the H0 late time direct measurements and the impact on the Dark Energy sector},
   volume={502},
   ISSN={1365-2966},
   url={http://dx.doi.org/10.1093/mnras/stab187},
   DOI={10.1093/mnras/stab187},
   number={2},
   journal={Monthly Notices of the Royal Astronomical Society},
   publisher={Oxford University Press (OUP)},
   author={Di Valentino, Eleonora},
   year={2021},
   month=jan, pages={2065–2073} }

@misc{poulin_2023,
      title={The Ups and Downs of Early Dark Energy solutions to the Hubble tension: a review of models, hints and constraints circa 2023}, 
      author={Vivian Poulin and Tristan L. Smith and Tanvi Karwal},
      year={2023},
      eprint={2302.09032},
      archivePrefix={arXiv},
      primaryClass={astro-ph.CO},
      url={https://arxiv.org/abs/2302.09032}, 
}

@ARTICLE{2022JHEAp..34...49A,
       author = {{Abdalla}, Elcio and {Abell{\'a}n}, Guillermo Franco and {Aboubrahim}, Amin and {Agnello}, Adriano and {Akarsu}, {\"O}zg{\"u}r and {Akrami}, Yashar and {Alestas}, George and {Aloni}, Daniel and {Amendola}, Luca and {Anchordoqui}, Luis A. and {Anderson}, Richard I. and {Arendse}, Nikki and {Asgari}, Marika and {Ballardini}, Mario and {Barger}, Vernon and {Basilakos}, Spyros and {Batista}, Ronaldo C. and {Battistelli}, Elia S. and {Battye}, Richard and {Benetti}, Micol and {Benisty}, David and {Berlin}, Asher and {de Bernardis}, Paolo and {Berti}, Emanuele and {Bidenko}, Bohdan and {Birrer}, Simon and {Blakeslee}, John P. and {Boddy}, Kimberly K. and {Bom}, Clecio R. and {Bonilla}, Alexander and {Borghi}, Nicola and {Bouchet}, Fran{\c{c}}ois R. and {Braglia}, Matteo and {Buchert}, Thomas and {Buckley-Geer}, Elizabeth and {Calabrese}, Erminia and {Caldwell}, Robert R. and {Camarena}, David and {Capozziello}, Salvatore and {Casertano}, Stefano and {Chen}, Geoff C.-F. and {Chluba}, Jens and {Chen}, Angela and {Chen}, Hsin-Yu and {Chudaykin}, Anton and {Cicoli}, Michele and {Copi}, Craig J. and {Courbin}, Fred and {Cyr-Racine}, Francis-Yan and {Czerny}, Bo{\.z}ena and {Dainotti}, Maria and {D'Amico}, Guido and {Davis}, Anne-Christine and {de Cruz P{\'e}rez}, Javier and {de Haro}, Jaume and {Delabrouille}, Jacques and {Denton}, Peter B. and {Dhawan}, Suhail and {Dienes}, Keith R. and {Di Valentino}, Eleonora and {Du}, Pu and {Eckert}, Dominique and {Escamilla-Rivera}, Celia and {Fert{\'e}}, Agn{\`e}s and {Finelli}, Fabio and {Fosalba}, Pablo and {Freedman}, Wendy L. and {Frusciante}, Noemi and {Gazta{\~n}aga}, Enrique and {Giar{\`e}}, William and {Giusarma}, Elena and {G{\'o}mez-Valent}, Adri{\`a} and {Handley}, Will and {Harrison}, Ian and {Hart}, Luke and {Hazra}, Dhiraj Kumar and {Heavens}, Alan and {Heinesen}, Asta and {Hildebrandt}, Hendrik and {Hill}, J. Colin and {Hogg}, Natalie B. and {Holz}, Daniel E. and {Hooper}, Deanna C. and {Hosseininejad}, Nikoo and {Huterer}, Dragan and {Ishak}, Mustapha and {Ivanov}, Mikhail M. and {Jaffe}, Andrew H. and {Jang}, In Sung and {Jedamzik}, Karsten and {Jimenez}, Raul and {Joseph}, Melissa and {Joudaki}, Shahab and {Kamionkowski}, Marc and {Karwal}, Tanvi and {Kazantzidis}, Lavrentios and {Keeley}, Ryan E. and {Klasen}, Michael and {Komatsu}, Eiichiro and {Koopmans}, L{\'e}on V.~E. and {Kumar}, Suresh and {Lamagna}, Luca and {Lazkoz}, Ruth and {Lee}, Chung-Chi and {Lesgourgues}, Julien and {Levi Said}, Jackson and {Lewis}, Tiffany R. and {L'Huillier}, Benjamin and {Lucca}, Matteo and {Maartens}, Roy and {Macri}, Lucas M. and {Marfatia}, Danny and {Marra}, Valerio and {Martins}, Carlos J.~A.~P. and {Masi}, Silvia and {Matarrese}, Sabino and {Mazumdar}, Arindam and {Melchiorri}, Alessandro and {Mena}, Olga and {Mersini-Houghton}, Laura and {Mertens}, James and {Milakovi{\'c}}, Dinko and {Minami}, Yuto and {Miranda}, Vivian and {Moreno-Pulido}, Cristian and {Moresco}, Michele and {Mota}, David F. and {Mottola}, Emil and {Mozzon}, Simone and {Muir}, Jessica and {Mukherjee}, Ankan and {Mukherjee}, Suvodip and {Naselsky}, Pavel and {Nath}, Pran and {Nesseris}, Savvas and {Niedermann}, Florian and {Notari}, Alessio and {Nunes}, Rafael C. and {{\'O} Colg{\'a}in}, Eoin and {Owens}, Kayla A. and {{\"O}z{\"u}lker}, Emre and {Pace}, Francesco and {Paliathanasis}, Andronikos and {Palmese}, Antonella and {Pan}, Supriya and {Paoletti}, Daniela and {Perez Bergliaffa}, Santiago E. and {Perivolaropoulos}, Leandros and {Pesce}, Dominic W. and {Pettorino}, Valeria and {Philcox}, Oliver H.~E. and {Pogosian}, Levon and {Poulin}, Vivian and {Poulot}, Gaspard and {Raveri}, Marco and {Reid}, Mark J. and {Renzi}, Fabrizio and {Riess}, Adam G. and {Sabla}, Vivian I. and {Salucci}, Paolo and {Salzano}, Vincenzo and {Saridakis}, Emmanuel N. and {Sathyaprakash}, Bangalore S. and {Schmaltz}, Martin and {Sch{\"o}neberg}, Nils and {Scolnic}, Dan and {Sen}, Anjan A. and {Sehgal}, Neelima and {Shafieloo}, Arman and {Sheikh-Jabbari}, M.~M. and {Silk}, Joseph and {Silvestri}, Alessandra and {Skara}, Foteini and {Sloth}, Martin S. and {Soares-Santos}, Marcelle and {Sol{\`a} Peracaula}, Joan and {Songsheng}, Yu-Yang and {Soriano}, Jorge F. and {Staicova}, Denitsa and {Starkman}, Glenn D. and {Szapudi}, Istv{\'a}n and {Teixeira}, Elsa M. and {Thomas}, Brooks and {Treu}, Tommaso and {Trott}, Emery and {van de Bruck}, Carsten and {Vazquez}, J. Alberto and {Verde}, Licia and {Visinelli}, Luca and {Wang}, Deng and {Wang}, Jian-Min and {Wang}, Shao-Jiang and {Watkins}, Richard and {Watson}, Scott and {Webb}, John K. and {Weiner}, Neal and {Weltman}, Amanda and {Witte}, Samuel J. and {Wojtak}, Rados{\l}aw and {Yadav}, Anil Kumar},
        title = "{Cosmology intertwined: A review of the particle physics, astrophysics, and cosmology associated with the cosmological tensions and anomalies}",
      journal = {Journal of High Energy Astrophysics},
         year = 2022,
        month = jun,
       volume = {34},
        pages = {49-211},
          doi = {10.1016/j.jheap.2022.04.002},
archivePrefix = {arXiv},
       eprint = {2203.06142},
 primaryClass = {astro-ph.CO},
       adsurl = {https://ui.adsabs.harvard.edu/abs/2022JHEAp..34...49A}
}

@ARTICLE{Halo_Galaxy_connect,
   title={The Connection Between Galaxies and Their Dark Matter Halos},
   volume={56},
   ISSN={1545-4282},
   url={http://dx.doi.org/10.1146/annurev-astro-081817-051756},
   DOI={10.1146/annurev-astro-081817-051756},
   number={1},
   journal={Annual Review of Astronomy and Astrophysics},
   publisher={Annual Reviews},
   author={Wechsler, Risa H. and Tinker, Jeremy L.},
   year={2018},
   month=sep, pages={435–487} }

@ARTICLE{Knox_2020,
   title={Hubble constant hunter’s guide},
   volume={101},
   ISSN={2470-0029},
   url={http://dx.doi.org/10.1103/PhysRevD.101.043533},
   DOI={10.1103/physrevd.101.043533},
   number={4},
   journal={Physical Review D},
   publisher={American Physical Society (APS)},
   author={Knox, L. and Millea, M.},
   year={2020},
   month=Feb }

@ARTICLE{Heitmann_2019,
   title={HACC Cosmological Simulations: First Data Release},
   volume={244},
   ISSN={1538-4365},
   url={http://dx.doi.org/10.3847/1538-4365/ab3724},
   DOI={10.3847/1538-4365/ab3724},
   number={1},
   journal={The Astrophysical Journal Supplement Series},
   publisher={American Astronomical Society},
   author={Heitmann, Katrin and Uram, Thomas D. and Finkel, Hal and Frontiere, Nicholas and Habib, Salman and Pope, Adrian and Rangel, Esteban and Hollowed, Joseph and Korytov, Danila and Larsen, Patricia and Allen, Benjamin S. and Chard, Kyle and Foster, Ian},
   year={2019},
   month=Sep, pages={17} }

@ARTICLE{Riechers_2022,
   title={Microwave background temperature at a redshift of 6.34 from H2O absorption},
   volume={602},
   ISSN={1476-4687},
   url={http://dx.doi.org/10.1038/s41586-021-04294-5},
   DOI={10.1038/s41586-021-04294-5},
   number={7895},
   journal={Nature},
   publisher={Springer Science and Business Media LLC},
   author={Riechers, Dominik A. and Weiss, Axel and Walter, Fabian and Carilli, Christopher L. and Cox, Pierre and Decarli, Roberto and Neri, Roberto},
   year={2022},
   month=Feb, pages={58–62} }

@ARTICLE{Press_and_Schechter,
       author = {{Press}, William H. and {Schechter}, Paul},
        title = "{Formation of Galaxies and Clusters of Galaxies by Self-Similar Gravitational Condensation}",
      journal = {\apj},
         year = 1974,
        month = feb,
       volume = {187},
        pages = {425-438},
          doi = {10.1086/152650},
       adsurl = {https://ui.adsabs.harvard.edu/abs/1974ApJ...187..425P}
}

@ARTICLE{Davis_1985,
       author = {{Davis}, M. and {Efstathiou}, G. and {Frenk}, C.~S. and {White}, S.~D.~M.},
        title = "{The evolution of large-scale structure in a universe dominated by cold dark matter}",
      journal = {\apj},
         year = 1985,
        month = may,
       volume = {292},
        pages = {371-394},
          doi = {10.1086/163168},
       adsurl = {https://ui.adsabs.harvard.edu/abs/1985ApJ...292..371D}
}

@ARTICLE{Jarvis_2004,
   title={The skewness of the aperture mass statistic},
   volume={352},
   ISSN={1365-2966},
   url={http://dx.doi.org/10.1111/j.1365-2966.2004.07926.x},
   DOI={10.1111/j.1365-2966.2004.07926.x},
   number={1},
   journal={Monthly Notices of the Royal Astronomical Society},
   publisher={Oxford University Press (OUP)},
   author={Jarvis, M. and Bernstein, G. and Jain, B.},
   year={2004},
   month=Jul, pages={338–352} }

@ARTICLE{Zeldovich_1970,
       author = {{Zel'dovich}, Ya. B.},
        title = "{Gravitational instability: An approximate theory for large density perturbations.}",
      journal = {\aap},
         year = 1970,
        month = mar,
       volume = {5},
        pages = {84-89},
       adsurl = {https://ui.adsabs.harvard.edu/abs/1970A&A.....5...84Z}
}

@ARTICLE{Cai_2026,
   title={The Hubble Tension: A Decade Review},
   volume={26},
   ISSN={2397-6209},
   url={http://dx.doi.org/10.1088/1674-4527/ae842f},
   DOI={10.1088/1674-4527/ae842f},
   number={8},
   journal={Research in Astronomy and Astrophysics},
   publisher={IOP Publishing},
   author={Cai, Rong-Gen and Wang, Shao-Jiang},
   year={2026},
   month=Jul, pages={084011} }

@ARTICLE{2026PhRvD.113f3519P,
       author = {{Poulin}, Vivian and {Smith}, Tristan L. and {Calder{\'o}n}, Rodrigo and {Simon}, Th{\'e}o},
        title = "{Impact of ACT DR6 and DESI DR2 for early dark energy and the Hubble tension}",
      journal = {\prd},
         year = 2026,
        month = mar,
       volume = {113},
       number = {6},
          eid = {063519},
        pages = {063519},
          doi = {10.1103/bx25-1g5d},
archivePrefix = {arXiv},
       eprint = {2505.08051},
 primaryClass = {astro-ph.CO},
       adsurl = {https://ui.adsabs.harvard.edu/abs/2026PhRvD.113f3519P}
}

@misc{freedman2025statusreportchicagocarnegiehubble,
      title={Status Report on the Chicago-Carnegie Hubble Program (CCHP): Measurement of the Hubble Constant Using the Hubble and James Webb Space Telescopes}, 
      author={Wendy L. Freedman and Barry F. Madore and In Sung Jang and Taylor J. Hoyt and Abigail J. Lee and Kayla A. Owens},
      year={2025},
      eprint={2408.06153},
      archivePrefix={arXiv},
      primaryClass={astro-ph.CO},
      url={https://arxiv.org/abs/2408.06153}, 
}

@article{Hoyt_2026,
   title={The Chicago Carnegie Hubble Program: Improving the Calibration of Type Ia Supernovae with JWST Measurements of the Tip of the Red Giant Branch},
   volume={1002},
   ISSN={1538-4357},
   url={http://dx.doi.org/10.3847/1538-4357/ae29eb},
   DOI={10.3847/1538-4357/ae29eb},
   number={1},
   journal={The Astrophysical Journal},
   publisher={American Astronomical Society},
   author={Hoyt, Taylor J. and Jang, In Sung and Freedman, Wendy L. and Madore, Barry F. and Owens, Kayla A. and Lee, Abigail J.},
   year={2026},
   month=Apr, pages={11} }

@ARTICLE{EDE_galaxies_cosmic_rush,
   title={The Cosmic Rush Hour: Rapid formation of bright, massive, disky, star-forming galaxies as signatures of early-universe physics},
   volume={550},
   ISSN={1365-2966},
   url={http://dx.doi.org/10.1093/mnras/stag1227},
   DOI={10.1093/mnras/stag1227},
   number={2},
   journal={Monthly Notices of the Royal Astronomical Society},
   publisher={Oxford University Press (OUP)},
   author={Shen, Xuejian and Zier, Oliver and Vogelsberger, Mark and Boylan-Kolchin, Michael and Hernquist, Lars and Tacchella, Sandro and Naidu, Rohan P},
   year={2026},
   month=Jun }
\end{document}